\documentclass[twocolumn,showpacs,preprintnumbers,amsmath,amssymb]{revtex4}
\usepackage{graphicx} 
\usepackage{dcolumn}
\usepackage{bm}
   
\begin{document}


\title{
The unrestricted Skyrme-tensor  time-dependent Hartree-Fock \\
and its application to the nuclear response 
 from spherical to triaxial nuclei} 
\author{S.~Fracasso}
\email{S.Fracasso@surrey.ac.uk}
\author{E.B.~Suckling}
\author{P.D.~Stevenson}
 \affiliation{
Department of Physics, University of Surrey, Guildford, Surrey GU2 7XH, United 
 Kingdom\\
}

\date{\today}

\begin{abstract}
 The nuclear time-dependent Hartree-Fock model formulated in the three-dimensional space, 
 based on the full standard Skyrme  energy density functional complemented with the tensor force, 
 is presented  for the first time. Full self-consistency is achieved by the model.  
 The application to the isovector  giant dipole resonance  is discussed in the linear limit, 
  ranging from spherical nuclei ($^{16}$O,$^{120}$Sn) 
 to systems displaying  axial or triaxial deformation  ($^{24}$Mg, $^{28}$Si, $^{178}$Os, $^{190}$W ,$^{238}$U).  
 \\  Particular attention   is paid to the spin-dependent terms from the central sector of the functional, recently included together with the tensor. 
  They turn out  to be capable of producing a qualitative change on the strength distribution in this channel. The effect on the deformation properties is also discussed.  
   The quantitative effects  on the linear response are small and, overall, the giant dipole energy remains unaffected. 
\\  Calculations are compared to  predictions from the (quasi)-particle random phase approximation  
  and  experimental data    where available, 
  finding good agreement. 

 \end{abstract}


\maketitle

 \section{Introduction}
This work presents the  full implementation
of the  Skyrme energy density functional (EDF), including the tensor terms,   
  in time-dependent Hartree-Fock (TDHF)~\cite{PRS}. The TDHF equations have been formulated 
 on a three dimensional (3D) Cartesian grid with no spatial symmetry restrictions and include all time-odd terms. Consequently, it is possible 
to study  different phenomena in both intrinsically spherical and deformed systems,  in particular triaxiality, 
within the same formalism and without neglecting terms in the particle-hole 
(p-h) channel of the nuclear (and Coulomb) interaction.  
 They can range from nuclear 
 structure features like giant resonances (including nonlinearities 
  and decay by particle emission)  and large amplitude dynamics like collisions of nuclei (including fusion and fission processes).   
 In this paper we focus on the first application, while another one about  nucleus-nucleus collisions~\cite{ES} will follow 
as an extension of recent related work~\cite{ES1}-\cite{Iwata}. \\
\indent
 The tensor force included in the original description of the 
 Skyrme interaction~\cite{Skyrme_old} had mostly been omitted from Hartree-Fock calculations after initial explorations~\cite{Stancu}.  It has been the object of considerable
 interest in the past few years, when particular effort has been spent to attest the real limits of the mean-field approach. 
 For a review of the ``new generation'' of tensor studies 
 we refer to~\cite{Les}, the authors of which  extensively analysed the performance of  several Skyrme-tensor functionals  on ground-state properties of spherical systems.  In such cases,  only the so-called time-even terms of the EDF, 
   built on densities which maintain the same sign under the time-reversal operation, 
      contribute. 
  Their coupling constants   are typically  fixed 
 through a selection of  experimental and empirical 
  data.  \\
   A situation similar to the spherical case, although not identical,  occurs in 
   axially deformed nuclei. The full  spatial rotational invariance 
 is lost in the  intrinsic system of reference,  
giving rise to  extra contributions to the one-body potential with respect to the 
spherical case~\cite{BenRMP}; however,  
 one expects the parity and the total angular momentum projection along the symmetry axis 
 are still good quantum numbers,  
   the time-reversal invariance is  preserved and 
 all the 
 time-odd terms of the functional 
   vanish in the ground-state. This fact is  maintained when 
 pairing effects are introduced, as long as the particle-particle scattering involves pairs in time-reversal states. 
   In Ref.~\cite{Ben-Les}, 
 the performance of  time-even tensor terms 
 fixed on  spherical systems~\cite{Les} 
 was studied against  deformed ground-state 
properties.
  Fewer analyses are 
 available for triaxial ground-states (cf.~\cite{Guzman}-\cite{Robledo} and 
 references therein),  where  
 the spatial symmetry properties are further broken,  
 but the time-reversal invariance 
 is known to be still preserved.  
\\
 Increasing the complexity of the problem, a  different situation  
 occurs in odd-A and odd-odd nuclei and, in general, in the presence of cranking, when  the time-reversal invariance is broken and the Kramer's degeneracy is 
 removed~\cite{Schunk}-\cite{Minkov}.  In such situations, in fact, time-odd terms become active. 
 The effects from the time-odd  terms  that originate from the Skyrme 
   central and   spin-orbit  force 
 have been studied through their influence on rotational bands of superdeformed nuclei~\cite{Dob}.  More recently, the full Skyrme 
  time-odd sector, 
 including the tensor terms, has been taken into account too~\cite{VHPRC}. 
\\\indent 
 Regarding vibrational states,  
 on the side of the (non-relativistic)  
 self-consistent random phase approximation (RPA), which does not model the pairing correlations and 
 which corresponds to the 
 small  amplitude (linear) limit 
 of TDHF,  progress has been obtained by including the residual Skyrme-tensor interaction in  formulations  on a  spherical p-h configuration basis. 
  As known,  the time-odd terms play an essential role in explaining the nuclear response in some cases.  
 Calculations including the tensor 
  in both some ``neutral''~\cite{Cao} and 
 charge-exchange channels~\cite{Bai} have been published.  
  Similarly to previous work, 
 the relative contribution of the residual two-body  versus the underlying  mean-field tensor effects was discussed by these authors, but no further 
 considerations on the various functional's terms and their interplay was made.    Skyrme-tensor RPA calculations in infinite matter~\cite{Davesne} have been performed as well.
\\\indent 
The preparation of this work has been also motivated 
 by the lack of a 
  self-consistent implementation 
 of the Skyrme functional complemented by the tensor interaction in 
 a deformed (quasi-)particle random-phase approximation 
  ((Q)RPA), 
  as the more general TDHF  approach is capable to fill such a gap,  
by allowing more applications on the same footing.   
 In this TDHF model all the Skyrme-tensor EDF terms have been derived and implemented;  
 the specific role played by some of them, where the focus is on spin-dependent terms that 
 were previously neglected,  will be outlined 
  and issues related to self-consistency discussed. 
  Although the comparison between TDHF and RPA approaches 
  is not the aim  of this paper 
  and more  can be found elsewhere, 
 in  section~\ref{comp} we will recall some 
       (standard)  terminology and  elements useful  to  follow the discussion. \\ \indent
 Previous 
 applications  of TDHF to zero temperature giant resonances,    
 including  time-odd terms from modern 
 Skyrme EDFs  and exploiting  
 computational power which was  not available in the past,  
 exist. 
 In Ref.~\cite{Nakatsukasa}, 
  the linear response in some light spherical and 
 deformed nuclei was studied by 
 comparing different approaches (TDHF with absorbing 
 boundary conditions and continuum RPA formulated with the 
 Green's functions formalism); 
  the terms depending 
 on the square of the momentum density ($\bm j^2$), which affects the effective mass, and on the square of the spin density ($\bm S^2$) were discussed, 
 while other spin-dependent terms were omitted from those calculations.  
 A more complete implementation, still without the tensor terms, 
    was employed to model $^{16}$O-$^{16}$O collisions in Ref.~\cite{Umar2006}. 
   The authors of Ref.~\cite{MAR}  pointed out the relevance of the Skyrme time-odd spin-orbit in  
  suppressing spurious  spin excitations in free translational motion, 
  associated to a nonphysical energy dissipation in heavy-ion collisions. 
   This was explained as a consequence of the  Galilean invariance restoration  when  
  the spin-orbit sector of the functional is fully included. 
  \\\indent 
After the formalism is introduced (Sec.~\ref{th}), we present (Sec.~\ref{res}) the results about the 
isovector dipole  response (IVD) for some 
 representative cases. 
 It is  well known that most of the IVD transition strength  is concentrated in the giant dipole resonance (IVGDR or simply GDR), the first 
  collective nuclear excitation   that was discovered~\cite{BG}-\cite{GoldT}.
   Although the major effect from the inclusion of the spin-dependent terms  
 is  not expected in this case, 
 we focus on it 
 according to the tradition of the TDHF calculations, as this is the simplest, experimentally 
 well known,  
  nonspherical excitation 
 that can be reproduced. 
  Many theoretical simulations are available, ranging from  more phenomenologic  approaches 
(where the important  
 dependence on external input can produce accurate calculations, but drastically limits 
 the predictive power),
 down  to microscopic descriptions with no \emph{ad-hoc} 
 adjustable parameters, 
 to which group our model belongs. \\ 
  In Sec.~\ref{Concl} conclusions will be summarised. 
  The derivation of the tensor contribution to the Skyrme EDF 
 is provided in App.~\ref{App1} as an extension of~\cite{Engel}, 
  where the functional form of the 
 energy density from the central and spin-orbit part of the  force 
   was written without assuming a  time-reversal invariant system. Finally, 
  the  expressions and  possible  implementations of the densities and currents of the functional have been added 
 in App.~\ref{App2}.  

\section{Formalism}
\label{th}
 The first part of this section recalls 
the basic features of the TDHF theory and the numerics adopted in this work, 
   while the second one   focuses on the full Skyrme-tensor energy density functional, implemented in TDHF with no symmetry restrictions. 
 Some of the performed verification and validation 
  tests  will be discussed.  

\subsection{Time-dependent Hartree-Fock model}
\label{TDHF}
 As mentioned in the Introduction, the theoretical framework chosen to implement the unrestricted Skyrme energy density functional, to which the following section is dedicated,   
 is the 3D Cartesian  time-dependent Hartree-Fock.  
 In this theory~\cite{Dirac}, 
  with semi-classic limit given by the Vlasov equation, 
  well known in plasma physics and 
  astrophysics,  
 the interaction among particles  is modelled in terms 
 of the interaction with the one-body potential, 
 which, after (or under) the action of an external field,  
 changes in time through the dependence on the 
  density itself.  The master expression of the theory derives from  
   the Von Neumann equation, the quantum-mechanical version  of the classical Liouville equation which 
 provides the 
 time evolution of the A-body density matrix. 
 Through the constraint $\hat\rho_A^2=\hat\rho_A$ for the full density matrix, the system can be represented by a pure state, solution of the  
 associated Schroedinger equation. In TDHF, 
  the previous relation is assumed to hold for the one-body density matrix  $\hat\rho_1$ 
(hereafter $\hat\rho$), so that the A-particle 
wave-function can be represented by a Slater determinant at any time (the backward relation holds too;  
 details can be found in Ref.~\cite{Koonin_thesis}).   
 As a matter of fact, 
the Von Neumann equation, which gives rise to the known 
 BBGKY hierarchy 
 for the reduced 1-body, 2-body, etc. densities, 
 now 
 simplifies to a one-body problem  
\begin{equation}
i\hbar \delta_t\hat \rho=\left[ \hat h,\hat \rho \right],    
\label{rho_ev}       
 \end{equation}  
which can be equivalently represented by 
 the (nonlinear) set 
 of one-body equations 
\begin{equation}
i\hbar \delta_t \psi^{(i)}(\bm x,t)= \hat h[\rho(t)] \psi^{(i)}(\bm x,t),  
\label{time_eq}  
\end{equation} 
 one for each spinor $\psi^{(i)}(\bm x,t)=\sum_{\omega=2m_s}
\phi^{(i)}(\bm x,\omega,t)\xi(\omega)$ identified by  the index $i$ 
($\omega=2m_s=\pm 1$),   
 with  initial conditions  given further  on (the dependence of the one-body Hamiltonian on the particle density can be generalised). 
  A time-dependent  external field can be inserted in the previous equation~(\ref{time_eq}). In this work, 
     it  will be  assumed  to simply act as a $\delta$ function in time,  
     used   to prepare the initial conditions for the time evolution.  
  The external field acting on the system can be defined, 
 in particular, as a one-body operator able to induce an isovector 
 response (protons and neutrons  move in opposition of phase) with no charge-mixing, that is in the form 
\begin{equation}
\hat F=\sum_{rs} <s|F|r>a^{\dagger}_ra_s, 
\end{equation}
 where 
 $a^+_r$ and $a_s$ are creation and annihilation operators in the HF basis and $F = \sum_i D f(\bm x_i, \omega_i) \tau_{z}(i)$.  
 This 
 external field  transfers energy (regulated by the strength $D$) and   
a selection of angular momentum 
to the nucleus, 
 causing a 
  displacement of the proton and neutron centres of mass, compatible deformation 
     and/or 
 spin fluctuations; 
 in such a way, the system is set in oscillation with all the possible frequencies. In other words, after the external action, the A-particle  wave-function is no longer an eigenstate of the static (``unperturbed'') Hamiltonian and starts to evolve in time.  In a microscopic picture, the 
  external kick induces  p-h transitions that are allowed by the selection rules,  
  so that, at time $t$=0,   
 each Hartree Fock eigenstate 
$\phi_i^{(0)}$  is 
 transformed 
 into a wave-packet 
\begin{equation}
\psi^{(i)}(\bm x,t=0)=e^{i\hat F} \phi_i^{(0)}(\bm x)=\sum_l \alpha_{il}\phi_l^{(0)}(\bm x), 
\label{init_cond}
\end{equation} 
where  the index $l$ spans the full  HF spectrum. 
 The label  $(i)$ at the left-hand side  of the previous equation simply enumerates the  wave-functions obtained by boosting the HF solution having a set 
 of  quantum numbers $i$.  
 Using~(\ref{init_cond}) as initial condition, where only the choice of the external operator $\hat F$ is arbitrary  and (partly) 
 defines the problem under study,  
 the equations~(\ref{time_eq}) are solved  
 and a new set of quasiparticles 
 $\{\psi^{(i)}(\bm x,t)\}$ 
is given at any time. 
Clearly, under the action of the one-body potential, mixings other than those initially generated by $\hat F$ are produced. 
 \\\indent 
 In practical implementations, according to long-standing  prescriptions~\cite{FK},  
 the time is discretised and 
  the 
 evolution operator 
 is implemented in a Taylor expansion. The time step size and the order at which 
 the expansion is arrested are two mutually dependent 
 parameters  chosen in order to ensure the total
 energy and norms are 
 conserved in the dynamic calculation within acceptable accuracy. \\
 The system response is analysed  by following the 
  time-evolution of the expectation value 
for the  one-body operator of interest  
 \begin{eqnarray} 
\langle  \hat O\rangle(t)
&=&\langle \Phi_t|\hat O|\Phi_t\rangle 
\label{resp} 
\end{eqnarray} 
(referred to the ground-state value).  
 One can 
  replace the A-particle Slater determinant  at time $t$ ($\Phi_t$)   
by the simple product of the involved 
   TDHF wave-functions  
 and recast 
  the previous equation as  
\begin{eqnarray} 
\langle \hat O\rangle(t)
&=&\sum_i \langle \psi^{(i)}(\bm x, t)|O(\bm x)|\psi^{(i)}(\bm x, t)\rangle,  
\label{respt}
\end{eqnarray}
 which can be conveniently  
 expressed in terms of 
   properly defined one-body densities. 
   For example, the well known (effective)  
  isovector dipole response along the $\lambda$ direction is based on the  
 expectation values  
\begin{eqnarray}
\langle \hat O_{1^-01}^{K} \rangle(t) 
&=& \frac{\tilde D}{A}\int \lambda \left(Z\rho_n(\bm x ,t)-N\rho_p(\bm x ,t)\right )
 d\bm x, 
\label{DIP}
\end{eqnarray}
  depending on  the time-dependent isovector density  
 $\rho_{10}(\bm x, t)=\rho_n(\bm x ,t)-\rho_p(\bm x ,t)$.  
 It is  associated to the choice 
 \begin{eqnarray}
f_{1K}(\bm x,\omega)
&=&\mathcal{Y}_{1K}(\bm x)|\bm x|I_{\sigma}, 
\label{opF}
\end{eqnarray}
for   $f(\bm x,\omega)$, 
where $I_{\sigma}$ is the identity operator in the spin space (kept implicit in what follows) and 
$\{\mathcal{Y}_{1K}\}_{K=-1,0,1}=\{
\sqrt{\frac{3}{4\pi}}
\frac{ \lambda}{|\bm x|}
\}_{\lambda=x,y,z}$  
is the set of $l=1$ 
real spherical harmonics 
written in Cartesian coordinates
($\tilde D=D\sqrt\frac{3}{4\pi}$, $e$=1). The subscript in Eq.~\ref{DIP} denotes the $J^{\pi}ST$ quantum numbers.  
   The signal obtained on top of a spherical ground-state  is clearly invariant under any rotation in space of the boost direction.  
 The  centre-of-mass correction in the definition of the operator is included in order to remove 
 the translational zero mode from the strength function (the boost needs proper  weigths as well).
\\
 Under the hypothesis   $\hat O=\hat F$, 
 it is possible to show  that the 
   transition strength distribution associated to $\hat O$ 
 \begin{equation} 
 S_{\hat O}(E)=\sum_{\nu}|\langle \nu|\hat O|-\rangle|^2 \delta (E-E{_\nu}),  
\label{SE1}
\end{equation}
where  $\nu$ labels the  excited states in the small amplitude limit and $|-\rangle$ is the HF vacuum, 
can be obtained 
 through  
\begin{equation}
 S_{\hat O}(E)=-\frac{1}{\pi}\frac{\mathcal{F}
[\langle  \hat O\rangle (t)]}{\mathcal{F} [D g(t)]},  
\label{SE2}
\end{equation}
 where $\mathcal{F}$ denotes the Fourier transform operation
 and $g(t)$ is the generalization  
  of the (factorized) time profile of the external operator. 
   The proof of the  equivalence between Eqs.~(\ref{SE1}) and~(\ref{SE2}) 
can be found in Refs.~\cite{Calv}-\cite{Stev} (the 
  rearrangement  arising from   density-dependent Hamiltonian terms is usually 
  considered negligible). 
\\\indent
 The presence of the factor $D$  in the denominator of~(\ref{SE2})  
  guarantees  that 
 the   response 
  is kept constant for  different boost strength in the small amplitude 
 (RPA)  regime. 
   This rescaling 
 can be used as an indicator to test the validity of the 
 assumption of linearity.  
  In fact, when the boost amplitude is increased above a certain 
  threshold, 
  the channel for  
  nonharmonic effects, which 
accounts for 3p-1h (3h-1p) vertices~\cite{sim_chom}
  in  a one-body view    is opened.  
\\
 Choices different from 
 $\hat O=\hat F$
   allow one 
 to select  the signal of  interest, 
 like a specific multipole or spin (isospin) component,  
 among the
 various response channels which can be opened by a given external operator 
 and  mixed up during the dynamic evolution.   
  For example, 
 the 1$^-$ states 
  are  a mixture 
 of both non-spin-flip ($S$=0, $S$=1) 
 and   spin-flip  ($S$=1) components, that is, 
 in a microscopic picture, nucleons can  also reverse the spin when undergoing particle-hole transitions.  
 A large fraction of the isovector $S$=1 dipole strength is collected in the giant spin-dipole resonance  
 (IVSDR or, hereafter,  SDR).  
   Several investigations on both the experimental and the 
 theoretical side (see \cite{SA} and references therein) have been dedicated to 
 the problem of the GDR-SDR splitting,   
  which markedly reflects the dependence of the effective interaction on the incident energy.  
A schematic model predicts the SDR to be lower in energy  than the GDR, 
 on the basis of a    residual interaction 
  being slightly less repulsive in the 
  spin-isospin ($S$=1, $T$=1) channel  
 than in the 
    isospin ($S$=0, $T$=1) one~\cite{Ost}.   
  From general arguments, in a limit case 
  one can find a highly collective 1$^{-}$ state exhausting the whole sum rule 
 for the dipole (or for the spin-dipole) 
operator, so that 
 zero transition probability  
   associated to the other one is left~\cite{Sag}.  
    In practice, this is less likely to occur,    
   the two resonances overlap and 
  1$^-$ states displaying not vanishing transition strength associated to either of the operators   
    can be 
  found to be degenerate or lying close in energy (see examples in Ref.~\cite{Wak}). 
\\\indent 
As giant resonances are above threshold, particle emission is expected and it can be modelled  in  
 TDHF~\cite{PRE}.  
A discussion about the lifetime of the TDHF wave-functions can be found, e.g., in Ref.~\cite{decay},  
 where  a comparison between the performance of 
 TDHF and the continuum-RPA  in the matter of escaping widths and the possibility 
 of comparing to experimental information was  discussed. \\
 Ref.~\cite{James} presented the effect of two-body correlations on the mass dispersion from  giant resonances, 
  incorporated in the standard TDHF as fluctuation of the one-body observable of interest on the basis of  
 the Balian-V\'en\'eroni approach~\cite{BV}, which turned out to be important. 
 As a matter of fact, 
 although TDHF takes into account the coupling to the continuum 
and   those anharmonicities that can be captured through one-body operators, 
 extensions are required 
  to fully describe the spreading width of the resonances.  
 Effort has been devoted to formulate extensions of TDHF
 which go beyond the (dynamic) 
 mean-field approximation,  
 capable of modeling 
  coherent collisional terms 
 as well as incoherent effects, 
   like those leading to 2p-2h admixtures into the strength function from   
 nondiagonal 
 couplings   to  low-lying 
  phonons 
   or  
 not collective 1p-1h bubbles~\cite{Lac_rev}.  
 We recall that   
 couplings of 1p-1h to 
  high-lying 2p-2h states 
 in  spin-isospin  
 channels 
 are  expected to be 
  favoured by the tensor,  
 a  mechanism which, 
 spreading  strength towards high energies, 
 has been invoked  
to fully explain 
  the quenching  of the Gamow-Teller (GTR) 
  and the 
   SDR (see Ref.~\cite{Ost}), 
   collective states  
  of particular interest for astrophysical processes and particle physics.  
  \\\indent 
If the standard TDHF approach and the extensions 
  recalled just above  
 describe a single path in the time domain,      
 stochastic formulations of the theory   
 are 
 available, 
 which were shown to be 
   closely related to the 
  Boltzmann-Langevin equation (cf.~\cite{Reinhh},~\cite{Reinh_Boltz} and references therein. 
  Cf.~also the discussion in Ref.~\cite{JamesTh}).

\subsubsection{Comments on TDHF and  RPA models} 
\label{comp} 
This section is not aimed at 
providing a complete review of similarities and differences between the two 
 approaches, 
 their various formulations or the 
  physics involved. 
 Instead, 
   some (standard) terminology is recalled, together with  
  a few elements useful to  better follow the discussion. 
  Besides the fact that TDHF is a more general theory than the RPA, 
  it is useful to remark on some technical differences when the former 
   is employed   in the linear limit.  
\\\indent 
The RPA framework, firstly formulated for plasma physics in 1953~\cite{BP}, involves, besides the construction of 
 the static mean-field,  the computation of the two-body  matrix elements of the residual interaction in the p-h channel.  
 In the most microscopic approaches, which make little use of 
 \emph{ad-hoc} adjustable parameters, 
  they are respectively defined as the first and 
 second derivative of the energy density functional (given by \emph{ansatz} 
 or built on  an effective interaction like Skyrme) with respect to the particle density fluctuation~\cite{War}.
  The  derivation must be performed from the  most general expression of the energy density, that is built  without the symmetry restrictions  
 allowed in special cases (cf.~Refs.~\cite{Vau},~\cite{Engel}; see also Ref.~\cite{ES}). 
   As a matter of fact, not only are ground-state spin saturation properties  
  (which here means  $\sum \psi_i^*(\bm x)\bm\sigma_i\psi_i(\bm x)=0$) 
    and spatial symmetries  usually broken when exciting the system, but also,  
  in some cases,  the excitation is mostly driven by terms that vanish in the  Hartree-Fock approximation.\\
 Self-consistency, as  intended  in the sense of the  RPA (we will not discuss   
 the case of the various types of separable RPAs; information can be found 
in Ref.~\cite{Nest})
 means  that EDF contributions are not dropped 
 when proceeding from the Hartree-Fock mean-field to the computation of excited states. Under this condition,  provided the 
 model space is complete, it is commonly known  that 
 symmetries  spontaneously broken by the Hartree-Fock approximation 
  are automatically restored and spurious modes,  within the limits of numerical accuracies, 
  become orthogonal to the physical spectrum and  degenerate with the ground-state (zero-modes)~\cite{Lane}. \\
  Concerning the  TDHF model 
  a full discussion about the symmetries breaking should be provided, 
 which is beyond the scope of this work.  We only add that,   
 as for the RPA approaches, numerical inaccuracies can produce spurious mixings and alter the strength distribution. 
  However,      a lack of ``consistency''  
 can be produced also in other ways, for example from an incomplete implementation of the functional at the mean-field level  
 with respect to  the procedure 
 adopted when fitting its  parameters, as well as  any 
   use which produces  lack or   overcounting of effects.  
  \\\indent 
 Regarding an effective two-body force with no density-dependent coupling constants, as for 
 the tensor terms by Skyrme, one can avoid  working out the energy density in 
 order to derive the mean-field and the 
  two-body residual interaction, because, in the absence of rearrangement, the latter would  be identical  to the starting force. However, 
  one  would miss, in this case, 
  the link of the interaction terms back to the corresponding  
  energy functional contribution,  which one usually refers to   
  when   comparing   different calculations. 
  Although it is recognized  that the globally fitted values of the Skyrme parameters should be interpreted by looking at the functional as a whole,  
 it is still interesting  
to study   the role of the various terms by setting some 
 coupling constants to zero, or by artificially modifying the functional in other ways.  
   If the consistency breakings produced in such a way 
      is not dramatic, this operation can reveal general  features  
 of the effective interaction and point to  
  drawbacks and credits of one specific parametrization.
 With such an aim, the use of Skyrme-like EDFs, that is   non-two-body interaction based EDFs, more or less rich in 
   density-dependent coupling constants and derivatives,   
  has become quite widely investigated. 
 \\
  In the standard TDHF theory, 
 only the mean-field operator 
  is required in both the static (HF ground-state) and dynamic (residual effects) 
 calculations.  The  implementation of the functional is, in this sense, much 
 less involved with respect to the 
  RPA approach and 
    the fullfilment of self-consistency 
   less demanding, expecially in comparison to 
    RPAs formulated in the coordinate space. 
  Clearly, the  effects corresponding to the various terms, coupled  one another through the densities, are not 
 expected to be simply additive.  It occurs in analogy to  RPA whenever   
  a different matrix is diagonalized once two-body contributions  are dropped. 
\\\indent 
Concerning the  microscopic understanding  of the results, 
  in the widely employed RPA  models formulated in   
  the  configuration  space  
 the equation of motion is reduced to the diagonalization of a matrix built on a p-h basis;  
   the solution immediately provides the microscopic structure of the excited states in terms of the simplest 
 excitations assumed 
  within the theory   and this repays the effort of building the two-body matrix elements.   
 The RPA transition strength distribution is built up by a linear combination of the 
 one-body p-h transition amplitudes (OBTAs) for the given operator, with weight depending on the expansion coefficients resulting from the diagonalization 
 (which reduce to 0 or 1 in the non-interacting limit, corresponding to the ``unperturbed'' response).  
 These ingredients provide together a useful  guideline to  microscopically interpret  the RPA response (at least in a weak-coupling regime). 
  One can understand, for example, which p-h wave-function contributes more strongly (in terms of OBTA and   
  relative weight) 
to an eigenstate displaying a relevant amount of transition probability for 
 the operator of interest. Moreover, one can attest 
   to  what extent the degree of mixing among the p-h states,  the mechanism 
   underpinning  the collectivity, where the relative importance between the strength of the residual interaction 
      and the unperturbed energies plays a key role, 
  is mainly due to one or the other term of the force.  
 The reader can refer to one example in Ref.~\cite{Comex}, which shows how  
 the residual Skyrme spin-orbit 
 is able to admix extra p-h configurations with opposite spins 
 in the considered low-lying RPA state.  \\
  Although less commonly studied, also considered some interpretation issues,  
    microscopic  information 
  can be extracted  from 
 TDHF simulations as well, provided the 
  requested operations 
 are built in. 
 Examples of microscopic analyses of TDHF calculations  can be 
 found in Refs.~\cite{Avez1}-\cite{Avez2}.  
  Further considerations on this topic are deferred to the future.

\subsection{Unrestricted Skyrme energy density functional} 
\label{sk}
   TDHF is a well established implementation of the 
  density functional approach  
in   nuclear physics~\cite{BenRMP}. 
 The starting point is the   
  static (effective) Hartree-Fock energy density $\mathcal {H}(\bm{x})$  
\begin{eqnarray}
 E_{HF}&=&\langle - |\hat H| - \rangle_{as}=
\int\mathcal{H}(\bm{x}) d\bm{x},   
\label{EHF}
\end{eqnarray}
 from which the 
  mean-field  
 can be derived. In our case, 
 in the previous equation  $\hat H$ is  the  effective Skyrme Hamiltonian 
  and $|-\rangle$ 
 is the Hartree-Fock ground-state. In general, a non-local energy density can 
 be defined. 
  Due to the effective nature of the nuclear Hamiltonian and the associated mean-field, the 
  Hartree-Fock (Hartree-like) equations 
 share the same philosophy of the electronic Kohn-Sham equations~\cite{KS}. \\
  Although the Skyrme interaction is constructed  as a contact force,  a fact that allows the implementation of the functional 
  in terms 
 of  one-body 
 densities depending on a single point of space, 
 finite range effects are  simulated through the derivatives of the 
 wave-functions. 
  In fact, 
  the standard Skyrme-tensor HF 
 energy density is expressed like a sum of terms bilinear in the particle or spin densities 
\begin{eqnarray}
\rho(\bm {x})&=&\rho(\bm x,\bm x')\Big|_{\bm{ x}=\bm{x}'}\label{dens1}\\
&=&
\sum_i\sum_{\omega}  \phi_i^{*
}(\bm{ x} ',\omega)\phi_i(\bm{ x},\omega) \Big|_{\bm{ x}=\bm {x}'}
\nonumber
\\
\bm{ S}(\bm{ x})&=&\bm S(\bm x,\bm x')\Big|_{\bm x=\bm x'}\label{dens2}\\
&=&\sum_i\sum_{\omega,\omega '} 
  \phi_i^{*
}(\bm{x} ',\omega')
\phi_i(\bm{ x},\omega) 
\langle 
\omega '|\bm{ \sigma}|\omega \rangle  \Big|_{\bm{ x}=\bm{x}'}\hspace{-0.7cm},\nonumber
\end{eqnarray}
 and   other kinds 
 of densities defined by  the up-to second order derivation of these 
  objects,   
  before the 
  local limit is taken (see Ref.~\cite{Carl} for higher order  EDFs).  
   In the equations above,  the  spin components 
  of the spinor $i$  are represented 
 by $|\omega\rangle$  (previously indicated by $\xi(\omega)$)
and  $\bm{ \sigma}$ are the Pauli matrices.   Once  the static densities are replaced by those   
built on  the solutions of the time-dependent one-body  
  equations, 
the total energy density 
 gains a time dependence 
 ($\mathcal{H}(\bm{x})\rightarrow \mathcal{H}(\bm{x,t})$), 
 although its functional form 
 remains unchanged with respect to the static case. 
\\\indent 
When only the central and spin-orbit terms of the Skyrme 
 force are retained, not all the combinations 
  allowed by the symmetries, among those depending on the second-order derivatives, 
   appear in the corresponding functional.  The tensor force provides a richer structure, 
 introducing in the $N-N$ interaction the dependence on the relative spin orientation  according  to 
\begin{eqnarray}
v_{\tau}(1,2)&=&V^{\tau}_{1,2}(\bm{ r}) \label{finite}\\
&=&4V^{\tau}_{1,2}(r)\left[ 
3\frac{(\bm{ \sigma}_1\cdot \bm{r})(\bm{ \sigma}_2\cdot \bm{ r})}{r^2}-\bm{ 
\sigma}_1\cdot \bm{\sigma}_2\right],  \nonumber
\end{eqnarray}
 where  $\bm{ r}$ is  the relative position of the two interacting particles  
  and  $V^{\tau}_{12}(r)$ is the 
 spatial form factor that  also  accounts for the isospin dependence. 
 It    is the only (local) force 
 able to explain  the  deuteron quadrupole moment, being   able to transfer to a two-particle wave-function 
 up to two units of orbital angular momentum  
 and  active only between spin triplet states.  
  As a one-body potential,  
 the tensor force is able to mix single-particle states that differ by one or two units  
 of orbital angular momentum, possibly introducing a parity 
 mixing.\\  
  Within a meson exchange model, the 
 form factor of Eq.~(\ref{finite}) is modelled as a Yukawa  potential and its zero range limit can be parametrized in the Skyrme form 
\begin{eqnarray}
v_{\tau}(1,2) 
&=&
 \frac{T}{2}\Big[(\bm{\sigma}_1 \cdot \bm{k}')(\bm{\sigma}_2 \cdot \bm{k}')\delta+
\delta
(\bm{\sigma}_1\cdot \bm{k})(\bm{\sigma}_2\cdot \bm{k})\Big]-\nonumber\\
&&\frac{T}{6}(\bm{\sigma}_1\cdot\bm{\sigma}_2)\Big[\bm{k}'^2\delta
+\delta
\bm{k}^2 \Big]+\nonumber\\
&&\label{vtens} 
\frac{U}{2}\Big[(\bm{\sigma}_1\cdot\bm{k}')\delta
(\bm{\sigma}_2\cdot\bm{k})+
(\bm{\sigma}_1\cdot\bm{k})\delta
(\bm{\sigma}_2\cdot\bm{k'})\Big ]-\nonumber\\ 
&&\frac{U}{6}(\bm{\sigma}_1\cdot\bm{\sigma}_2)\Big[\bm{k}'\cdot\delta
\bm{k}+
\bm{k'}\delta\cdot\bm{k}
\Big], 
\end{eqnarray}
where $\delta$ denotes 
$\delta(\bm{x}_1-\bm{x}_2)$ and, as usual, $\bm{k}=\frac{1}{2i}(
\bm{\nabla}
_1 -
\bm{\nabla}
_2 )$ 
and $\bm{k}'=-\frac{1}{2i}(
\bm{\nabla}
'_1-
\bm{\nabla}
'_2)$  
 respectively act on the right and on the left.
 This fully invariant interaction term  corresponds to the original one~\cite{Skyrme_old} 
 and  provides the same identical contribution to the Hartree-Fock
 energy density (App.~\ref{App1}) as the  more compact expression often employed in literature 
 (cf.~Ref.~\cite{Stancu} and more recent works). 
\\ 
By using basic angular momentum algebra, the previous equation can be 
 recast in terms of products of  tensors of rank 2 in spin and momentum (see eg. Ref.~\cite{Cao}).  
 \\  Although the tensor force is able to act between relative L=2 states, 
 where the centrifugal barrier keeps the nucleons apart  so that 
 the long range attractive part of the $N-N$ interaction ($r>3$ fm) is probed,
 there is currently no evidence that an intra-medium 
  finite-range 
    tensor force performs better than a contact, velocity-dependent, parametrization. 
 Some remarks about the quality of the zero range approximation for the tensor force  
 were proposed by the authors of 
Ref.~\cite{Brink07}. 
  Recent comparisons between Skyrme (SLy5+t of Ref.~\cite{PLB}) and, in particular, the GT2 Gogny force of Ref.~\cite{Ots}, 
   can be found in Ref.~\cite{MT}.  
  Not many  finite range effective forces complemented with the tensor  are available  and   further studies 
  are envisaged for  the future. 
\\\indent 
 In the proton-neutron formalism,  
the contribution to the energy density  associated to the tensor force~(\ref{vtens}), once the exchange is taken into account,  
reads  
\begin{widetext}
\begin{eqnarray}
\mathcal{H}_{tens}(\bm{x})&=&2B_{\nabla S}
\bm{\nabla}\cdot\bm{S_n}(\bm{x})\bm{\nabla}\cdot\bm{S_p}(\bm{x})+
A_{\nabla S}\sum_q\left(\bm{\nabla}\cdot\bm{S_q(\bm{x})}\right)^2+
2B_{J_0}J^0_n(\bm{x})J^0_p(\bm{x})+A_{J_0}\sum_q(J^0_q(\bm{x}))^2+\nonumber\\
&&2B_{J_1}\bm{J}_n(\bm{x})\cdot\bm{J}_p(\bm{x})+A_{J_1}\sum_q(\bm{J}_q(\bm{x}))^2+
2B_{J_2}\underline{J}_n(\bm{x})\underline{J}_p(\bm{x})+A_{J_2}\sum_q(\underline{J}_q(\bm{x}))^2+\nonumber\\
&&B_{F}\left[\bm{S}_n(\bm{x})\cdot\bm{F}_p(\bm{x})+\bm{S}_p(\bm{x})\cdot\bm{F}_n(\bm{x})\right]+A_F\sum_q\bm{S}_q(\bm{x})\cdot\bm{F}_q(\bm{x})+\nonumber\\
&&B_{G}\left[\bm{S}_n(\bm{x})\cdot\bm{G}_p(\bm{x})+\bm{S}_p(\bm{x})\cdot\bm{G}_n(\bm{x})\right]+A_G\sum_q\bm{S}_q(\bm{x})\cdot\bm{G}_q(\bm{x})+\nonumber\\
&&
B_{T}\left[\bm{S}_n(\bm{x})\cdot\bm{T}_p(\bm{x})+\bm{S}_p(\bm{x})\cdot\bm{T}_n(\bm{x})\right]+A_T\sum_q\bm{S}_q(\bm{x})\cdot\bm{T}_q(\bm{x})+\nonumber\\
&&
B_{\Delta S}(\bm{S}_n(\bm{x})\cdot\Delta\bm{S}_p(\bm{x})+\bm{S}_p(\bm{x})\cdot\Delta\bm{S}_n(\bm{x}))
+A_{\Delta S}\sum_q \bm{S}_q(\bm{x})\cdot\Delta\bm{S}_q(\bm{x}). 
\label{Htot}
\end{eqnarray}
\end{widetext}
\noindent
 For a generic term $\alpha$ of the functional, 
 the $A_{\alpha}$ and $B_{\alpha}$ coupling constants 
   enter the expression  of the proton or neutron mean-field,  
  respectively 
 felt by a particle with the same and opposite isospin. 
In terms  of the commonly employed   
isoscalar-isovector coefficients $C_T^{\alpha}$, 
 they satisfy the relations   
$A_{\alpha}=C_0^{\alpha}+C_1^{\alpha}$ and $B_{\alpha}=C_0^{\alpha}-C_1^{\alpha}$. 
 The $C_T^{\alpha}$ coupling constants are related to the Skyrme parameters as tabulated in Ref.~\cite{Perl}, according to  the 
 convention for the EDF explained below. 
 \\
 The most evident difference with respect to the standard  
 Skyrme functional,   
  based only on  the central and spin-orbit terms of the force,    
 is 
the presence of two pseudo-vector densities 
built on the tensor product of the nabla  operators 
\begin{eqnarray}
\hspace{-0.3cm}\bm F(\bm x)&=&\frac{1}{2}\left \{\left[ (\bm \nabla '\otimes \bm \nabla )  +(\bm \nabla \otimes \bm \nabla '    )\right]
   \bm S(\bm x,\bm x')  \right\}_{\bm x=\bm x'}\\
\hspace{-0.3cm}\bm G(\bm x)&=&\frac{1}{2}\left \{\left[ (\bm \nabla '\otimes \bm \nabla'   )+(\bm \nabla \otimes \bm \nabla    )\right]
   \bm S(\bm x,\bm x')  \right\}_{\bm x=\bm x'}  \label{GG}\end{eqnarray} 
where 
$F_{\mu}= 
\frac{1}{2}\sum_{\nu}[
 ( \bm \nabla '\otimes \bm \nabla )  
+( \bm \nabla \otimes \bm \nabla ')]_{\mu\nu}   
    S_{\nu}$ and $\otimes$ denotes the usual 
 tensor product between vectors $(\bm A\otimes \bm B)_{\mu\nu}=A_{\mu}B_{\nu}$. 
The $\bm S\cdot\bm F$ and $\bm S\cdot\bm G$ 
 terms are connected through the $(\bm\nabla\cdot \bm S)^2$ term according to the equation  
\begin{equation}
\int \left [\bm F(\bm x)+\bm G(\bm x)\right ]\cdot \bm S(\bm x) d\bm x=
-\frac{1}{2}\int  \left [\bm\nabla \cdot \bm S(\bm x)\right ]^2   d\bm x. 
\end{equation} 
   The previous relation can be used to successfully switch from~(\ref{Htot}) to the formulation of the EDF 
 of Ref.~\cite{Perl}, 
where the $\bm{G}$ 
density was not defined and only the $\bm{S}\cdot\bm{F}$ terms appear, 
 with  proper weights.  In this paper  we  will  work with the equivalent 
 formulation based on the choice $C_T^G=0$ too.   
   More details can be found in   App.~\ref{App1}, 
 where the derivation of Eq.~(\ref{Htot}) is provided.  
  Indeed, since the terms of the functional are not independent from each other, for example also the relation~\cite{Dob_Acta}   
\begin{equation} 
-\int \left [\bm\nabla \cdot \bm S(\bm x)\right ]^2d\bm x=\int \left[\bm S(\bm x)\bm\cdot \Delta\bm S(\bm x)+
\left [\bm\nabla \times \bm S(\bm x)\right ]^2 \right] d\bm x 
\end{equation}  
holds,  many equivalent formulations for the energy density can be  
 worked out. 
 \\
The $\bm S\cdot\bm F$ and $\bm S\cdot\bm G$ terms  contribute to mantain the Galilean invariant properties  of the standard Skyrme functional  
 when the tensor is made active.  
 In particular, 
 they 
 are related to the 
  pseudo-scalar, vector and pseudo-tensor 
 densities 
 $J_0, \bm J, \underline J$ 
 and the  pseudo-vector spin kinetic density $\bm{T}$. The latter is 
   defined by  
\begin{equation}
\bm T(\bm x)=\left[\bm\nabla\cdot\bm\nabla '\bm S(\bm x,\bm x')\right]_{\bm x=\bm x'} 
\end{equation} 
where a generic component $T_{\mu}$ depends on 
$ \bm\nabla\cdot\bm\nabla ' S_{\mu}$.
 The former three  objects are those that represent  the 
 trace, the antisymmetric and the symmetric parts~\cite{Les} 
of the pseudo-tensor spin current 
\begin{equation}
\stackrel{\leftrightarrow}J(\bm{x})=\frac{1}{2i}\left[(\bm\nabla-\bm\nabla')\otimes\bm{S}(\bm{x},\bm{x}')\right]_
{\bm{x}=\bm{x}'},
\end{equation}
which decomposes into  
\begin{equation}
J_{\mu\nu}=\frac{1}{3}\delta_{\mu\nu}J_0+\frac{1}{2}\sum_{i=x,y,z}
\epsilon_{i\mu\nu}\bm J_i+\underline J_{\mu\nu}, 
\end{equation}
 with  $\epsilon_{i\mu\nu}$  being the Levi-Civita tensor; 
  their square satisfy the relations 
\begin{eqnarray} 
J_0^2&=&\sum_{\mu\nu}J_{\mu\mu}J_{\nu\nu}\\
\bm J^2&=&\sum_{\mu\nu}J_{\mu\nu}\left(J_{\mu\nu}-J_{\nu\mu}\right)\\
\underline J^2&=&
\frac{1}{2}\sum_{\mu\nu}J_{\mu\nu}
\left(J_{\mu\nu}+J_{\nu\mu}\right)-\frac{1}{3}J_0^2
\label{eqJ}
\end{eqnarray}
and simple algebra leads to 
\begin{equation}
\stackrel{\leftrightarrow}J^2=\frac{1}{3}J_0^2+\frac{1}{2}
\bm J^2+\underline J^2;  
\end{equation}
the usual definition for the scalar product between tensors 
$\stackrel{\leftrightarrow}A\stackrel{\leftrightarrow}B=\sum_{\mu\nu}A_{\mu\nu}B_{\mu\nu}$  
has been used. 
 As long as triaxiality is absent, $J_0=Tr\stackrel{\leftrightarrow}J$ is zero (in the presence of subscripts, $J_0$ will be denoted 
  by $J^{(0)}$).  In spherical symmetric nuclei, the sum over $\mu , \nu$ in 
Eq.~(\ref{eqJ}) vanish as well~\cite{ES}, 
 so   
 one recovers the 
reduction of $\stackrel{\leftrightarrow}J^2$ to $\frac{1}{2}\bm J^2$. 
 The  $J_0, \bm J$ and $\underline J$ densities 
enter the central sector of the standard Skyrme functional with weights that allow 
 the replacement  
\begin{equation}
C^{J_0,c}_T
(J_T^{(0)})^2+C^{J_1,c}_T
\bm J_T^2+C^{J_2,c}_T\underline J_T^2=
C^{J,c}_T
\stackrel{\leftrightarrow}J_T^2
\end{equation}
for $T=0,1$, since  
$C^{J,c}_T=3C^{J_0,c}_T=2C^{J_1,c}_T=C^{J_2,c}_T$ ($C^{\alpha}_T=C_T^{\alpha,c}+C_T^{\alpha,t}$, where c=central, t=tensor).  
Unless  the tensor is switched on,   
the two implementations, in terms of 
$\stackrel{\leftrightarrow}J$~\cite{Engel}     
or the formulation where  $J_0$, $\bm J$ and $\underline J$ are kept separated~\cite{Perl},  
  are   interchangeable. 
 In this case, the Galilean invariance only relates the  spin-current terms
 to those depending on the spin kinetic  density $\bm{T}$  ($C_T^{T,c}=-C_T^{J_2,c}$).  
 When the tensor is introduced,  
  the following relations among the parameters must hold
  for the Galilean invariance to be respected 
\begin{eqnarray}
3C_T^{J_0}+C_T^T+2(C_T^F-C_T^G)&=&0\nonumber\\
4C_T^{J_1}+2C_T^T-(C_T^F-C_T^G)&=&0\label{GalGF}\\
2C_T^{J_2}+2C_T^T+(C_T^F-C_T^G)&=&0\nonumber 
\end{eqnarray} 
and $3/10C^{J_0,t}_T=-2/5C^{J_1,t}_T=C^{J_2,t}_T$.  
 This is realised  for the  Skyrme-tensor functional; 
 in order to prove the invariance, 
  it is useful to know  that  under the 
local Gauge (Galilean) operation  $T_G: \psi_i(\bm x)\rightarrow 
e^{i\bm k\bm x}\psi_i(\bm x)$, the $\bm{F}$ and 
 $\bm{G}$  
 densities  transform according to
\begin{eqnarray}
T_G(F_{\mu}(\bm x))&=&F_{\mu}(\bm x)+k_{\mu}\bm k\cdot \bm S(\bm x)+ k_{\mu}J_0(\bm x)+\nonumber \\
&&+\sum_{\nu}k_{\nu}J_{\mu\nu}(\bm x)\\
T_G(G_{\mu}(\bm x))&=&G_{\mu}(\bm x)-k_{\mu}\bm k\cdot \bm S(\bm x)- k_{\mu}J_0(\bm x)+\nonumber\\
&&-\sum_{\nu}k_{\nu}J_{\mu\nu}(\bm x).
\end{eqnarray}
 In total, the Galilean invariance imposes constraints 
 on 8 (6 without the tensor)  of the 14       
 time-odd coupling constants of the whole Skyrme-tensor functional 
 (in the isospin formalism 
 and in the formulation where $C_{T}^G$=0)  
 which, in such a way, can be determined through  the corresponding even partners. 
We refer to~\cite{Engel} 
(notice the misprint in  Eq.~(4.6) among the Galilean transformations, 
 where $\bm k^2$ has to be replaced by $\bm k^2\rho(\bm r)$)  
and Ref.~\cite{Perl} 
 for the left relations concerning the Galilean invariance properties 
and  we proceed with some other comments.  
\\\indent 
 The inclusion of the spin-current  
 $\stackrel{\leftrightarrow}J^2$ terms, which, as know, impact on the 
 spin-orbit splitting in the absence of spin saturation (when each pair of spin-orbit partners is not fully filled up), 
  is required in forces  that took them into account during the fit.  
  Neglecting them leads to  a breaking of consistency in terms of an 
 incomplete implementation of the functional.  
  For a more complete discussion   refer to~\cite{Fracasso07}, where
  a comparison between predictions from the SLy4 force, fitted without that contribution (``type I'' force),  and the  SLy5 set, 
 which included it (``type II''), 
was  considered; ambiguities 
 related to the former type of force were  pointed out. 
 In particular, it was discussed that 
 the contribution from the  $\stackrel{\leftrightarrow}J^2$ terms 
  in computing the excited states  
 can be more important  
  than in the  ground-state, 
   with the result that 
  standard fitting procedures, unable to capture some physics through the 
 ground-state observables,  
  can lead to poor results when computing 
    collective excitations.  
  For 
  this reason, 
   suppressing those terms from the residual interaction of    
   ``type I'' forces can lead to 
 an error larger than a RPA  
 self-consistency breaking approach, although (unwanted) case-by-case considerations 
 with such parametrizations might be required. 
\\
 It similarly holds for the time-odd sector of the functional. 
The 
 central $\bm{S}\cdot\bm{T}$ terms 
  have been often suppressed in the past 
 together with their Galilean partners 
 $\stackrel{\leftrightarrow}J^2$
 (cf.~e.g. Ref.~\cite{BenderGT}). 
   However, 
  the error associated to the suppression of 
 the $\bm{S}\cdot\bm{T}$ terms in finite systems   
  can be large.  
  In Ref.~\cite{Fracasso07} 
 it was shown that  
    the lack of these time-odd 
  terms  
 can  significantly alter  the energy location of the GTR, which 
 dominates the landscape 
 in the 
   (charge-exchange) $1^+$ channel. 
 It also turned out that 
  the GTR collectivity, which, as known,  experimentally  
  absorbs  the 60\% of the Ikeda-Fuji-Fujita sum rule, 
   can be badly affected  
 by relatively strong (negative) values of the $C^T_1$ 
 coupling constant, 
  a fact which explained the peculiar behaviour of the SLy5  
 parametrization    with respect to  other Skyrme forces.  
 Also,  relatively  too weak  $C_T^T$ values (which depend on the 
 $t_1$ and $t_2$ 
  Skyrme parameters)  
  would allow a too strong mixing 
  from the $\bm S^2$ terms (depending on $t_0$ and $t_3$, mainly fixed on bulk properties), 
 unless the proper balance, or alternative formulations, are found.  
   On such basis, one  conclusion  of Ref.~\cite{Fracasso07}  
 was that 
 the spin, velocity-dependent terms cannot simply be neglected  as a rule  
 and the parametrized EDFs 
 rather need improved  fitting procedures to model the 
 dependence on the  spin density, 
  possibly including (further)  constraints from collective properties.  \\
 Among the last developments in the central spin sector, 
  extensions of the $t_0$-$t_3$ spin-dependent terms  
  were  proposed in Ref.~\cite{Marg09}, 
   in connection to the  spin phase transitions that are known to be predicted by effective approaches 
  beyond the saturation density (cf.~Ref.~\cite{Rios} and references therein). 
\\\indent 
The last remark of this section concerns 
superfluidity. 
 An approximation still quite in use in order to  include pairing correlations in TDHF consists in 
 performing a BCS calculation at the mean-field level and evolving the dynamics on top of it. 
 The sums in Eqs.~(\ref{dens1}),~(\ref{dens2}) would 
 consequently span the larger set of states that defines the pairing window,  
 the expression of the 
 (``normal'') densities  gain nontrivial occupation factors   
  and the energy functional is complemented by terms depending on the densities' ``anomalous''  
 counterparts~\cite{Perl}.   \\
 A detailed description of nuclear 
 response  would require a TDHFB approach. 
  A full 3D-TDHFB 
 is a demanding computational task. 
 Some  of the most advanced works of this type  are represented 
   by Refs.~\cite{Ebata}-\cite{Stet} 
 and, for the Gogny case, by Ref.~\cite{Hascimoto}. 
  A full TDHFB implementation of the Skyrme functional, including also the tensor, is not 
 yet available. 
  The approximated and not fully self-consistent treatment of superfluidity 
  does  not affect our conclusions 
  on 
  the  p-h channel, although an extension which unifies the best efforts in both the p-h and p-p (particle-particle) channels is 
 envisaged for the future.

\subsubsection{Parameters, numerical tests and the centre-of-mass correction}
In this work  we use the  SLy5 force~\cite{Chabanat} with  
   tensor  parameters that were perturbatively 
defined in Ref.~\cite{PLB} 
and  employed in  subsequent applications starting with Ref.~\cite{Zou} (cf.~also Ref.~\cite{SucE} and references therein). 
 As  known,  the Lyon forces,  
   which, in particular, received constraints also from 
 microscopic calculations of the 
 neutron matter  equation of state,
  were tailored  to improve the isovector channel of the effective 
 interaction and thus the description of 
   exotic systems. 
 Like 
 other works, 
we also employ the force T44 of Ref.~\cite{Les},  
 characterized  
 by some different  features. For example, 
 the central+tensor isovector coupling constant 
 $C_1^{J_1,c+t}$ of T44 vanishes and, consequently, 
  one is left, in the spherical limit,  
 with the isoscalar tensor contribution only.  
   Moreover,  for all the terms mentioned in the previous section ($J^2_0$, 
 $\bm J^2$, $\underline{J}^2$, $\bm{S}\cdot\bm{T}$, $\bm{S}\cdot\bm{F}$), 
      the like ($C^T_0+C^T_1$) and unlike ($C^T_0-C^T_1$) particle tensor contributions to the mean-field operator have the same sign,  
   at variance with the SLy5 case where they are systematically opposite 
  each other (see Tab.~\ref{Tab1}).  
   \begin{table}
\caption{ Central and tensor coupling constant for the 
 $J_0^2$, $\bm{J}^2$, $\underline{J}^2$, $\bm{S}\cdot\bm{T}$,   
  $\bm{S}\cdot\bm{F}$ and    $\stackrel{\leftrightarrow}J^2$ 
 terms, 
 according to the isospin 
formulation 
 for the energy density functional.  All the values are 
 in MeV$\cdot$fm$^5$. \\
\label{Tab1} } 
\begin{ruledtabular}
\begin{tabular}{ccccc}
&\multicolumn{2}{c}{\text{SLy5}}&\multicolumn{2}{c}{\text{T44}}\\
	& \text{central (c)} & tensor (t) & central (c) & tensor (t)  \\ 			
\hline
$C_0^T$	& -15.67&14.&	  -59.01&	-24.40\\
$C_1^T$	&-64.53&54.&-52.03&	20.81 \\
$C_0^{J}$=$C_0^{J2}$&15.67&	7.	& 59.01&	-12.20	\\
$C_1^{J}$=$C_1^{J2}$&64.53&	27.&52.03&	10.41\\
$C_0^F$	&0&-42.& 0&	73.19	\\
$C_1^F$	&0&-162.&0& 	-62.43	\\
$C_0^{J0}$&5.22	&23.33	&19.67& 	 -40.66\\
$C_1^{J0}$&21.51&90.&17.34&  	34.69 	\\
$C_0^{J1}$&7.83	&-17.50& 29.50	&30.50\\
$C_1^{J1}$&32.27&-67.50&26.01&	-26.01\\
\end{tabular}
\end{ruledtabular}
\end{table} 
 The SLy5 and T44 forces  have strictly negative central  $C_T^T$  coefficients,  
  a quite common feature 
 of the standard Skyrme 
 sets (cf.~Tab.~II in Ref.~\cite{Cao2010}). 
  They are relatively strong for both cases 
 (with a different $C^T_0$/$C^T_1$ ratio),
 so  the effects from some of the newly implemented terms 
  are expected to be 
 emphasized with respect to other choices. 
    \begin{table}
\caption{ $A_T$ (first line) and $B_T$ (second line) coupling constants for the SLy5 force and, within parenthesis, 
  T44 when the tensor is switched off (``central'') or retained.    All the values are 
 in MeV$\cdot$fm$^5$.  
 \label{Tab2} }
\begin{ruledtabular}
\begin{tabular}{cccc}
               &	central (c)	&tensor (t) &	c+t \\	
\hline
$C_0^T$+$C_1^T$ &	-80.20 (-111.03)&	68.0 (-3.59)&	-12.20 (-114.62) \\	
$C_0^T$-$C_1^T$	& 48.87 (-6.98)&	-40.0 (-45.21)&	8.87	(-52.19	) \\
 \end{tabular}
\end{ruledtabular}
\end{table}
 With the inclusion of the tensor, 
   the $C^T_T$ coupling constants remain negative;   
  the SLy5 ones 
  receive a large reduction (90\% and 84\% for $C_0^T$ and $C_1^T$ respectively), while 
  the T44  $C_0^T+C_1^T$ and $C_0^T-C_1^T$ combinations  
    are both 
 strengthened in absolute value  
(see Tab.~\ref{Tab2} of this work).  
\\\indent 
 In the pairing sector, an (isovector) volume pairing force 
was implemented.  
For $^{120}$Sn we  set its (neutron-neutron) strength  to the value  
V$_0$(n)=243 MeV$\cdot$fm$^3$; in such a way, 
 the SLy5 calculations including the $\stackrel{\leftrightarrow}J^2$  terms well 
 reproduce ($<$1$\%$ of discrepancy) the experimental neutron pairing gap 
(1.32 MeV)   obtained with the standard  
  mass formula~\cite{dug}. 
 The associated energy cut-off is such that, as rule of thumb,  
  only one extra major shell is included in the BCS space. 
 In the following,  we will refer to 
the choice V$_0$(p)=V$_0$(n)=243 MeV$\cdot$fm$^3$ as  
``pairing set 1'' in order to distinguish it from other sets discussed later in connection to heavier nuclei. 
 In this work, 
  we do not stick to the isospin invariance 
 in the pairing channel adopted by some of us in the past~\cite{Fracasso05}  
 and we
 separately fix  the  proton and neutron   
  pairing strengths.  
\\\indent 
The model space in the static calculations is a
 cubic box of 24$^3$~fm$^3$  
 with the commonly employed 1 fm 
 grid step. 
    A one-half reduction of the grid spacing would reproduce 
 the HF single-particle spectrum with a better precision, causing 
   variations from tens to 100-200 keV on single-particle energies in $^{16}$O 
  (which do not affect the  features of the giant resonances we are interested in),  
    but it would importantly enlarge the computational 
 time of the dynamical simulation. 
 We consider the standard choice of 1 fm  
  to be a good trade-off for our purposes. 
 In particular, the convergence in the static was checked, as usual, 
  by looking at the stability of the single-particle energies. 
\\\indent 
  The dynamic calculations 
 were performed   
with reflecting boundaries   
over the same grid of  the static case, 
 but extending the box  side parallel to 
 the boost direction  
  to 64 fm for the lightest systems. 
In the other cases, 
 a smaller 
 squared box 
 was  used 
 to save computational time which, together with the other model space parameters, 
  had been employed in previous calculations~\cite{Rp}. 
 The box size, 
 the simulation time length 
 and the boost strength, chosen 
   small enough to stay in the linear regime, minimizing the particle 
 emission 
 but without  being dominated  by numerical noise, 
   must be considered  mutually dependent parameters, properly chosen 
  to avoid unwanted effects from particle reflection. The choice for the input parameters 
    can   also  be tested 
 by checking 
    the absence of  unphysical 
  fluctuations in the 
  particle number or against the stability of the response computed in a reduced volume 
 when rescaling the box size. 
  The tests were  successfully performed by looking at  the    heaviest nuclei. 
 \\The numerical accuracy was also  checked  by observing the free motion of a single ground-state nucleus on the grid, the typical 
 test for the Galilean invariance. 
The nucleus in the starting rest condition is boosted by an external force 
   and simply put into translation without, ideally,     
 internal excitation. 
 First, one notices that in the simple spin-saturated $^{16}$O,  
 the contribution to the total energy from the spin current $\stackrel{\leftrightarrow}J$ 
is expected to be zero, the same occurring 
 for the energy contributions associated to the time-odd terms. 
 Simulating the translation, 
 the former quantity turns out to be of the order of 10$^{-4}$ MeV and it 
 oscillates in time   
  at  the 
 10$^{-5}$ MeV level, 
 while 
 the HF single-particle energies change by a few tens of keV with respect to the ``basic'' calculation that
 neglects the $\stackrel{\leftrightarrow}J^2$ terms. 
  The most important error source is  due to the choice of the grid spacing size: 
   the accuracy  
 improves by one order of magnitude  
  after  a one-half reduction of this parameter. 
In order for the functional to be analytically Galilean invariant,
the inclusion of the contributions associated to the pseudo-tensor 
 $\stackrel{\leftrightarrow}J$ 
   requires, as said, the $\bm{S}\cdot\bm{T}$ ones too and, in the 
 presence of the tensor, 
 the $\bm{S}\cdot\bm{F}$ ($\bm{S}\cdot\bm{G}$) terms as well. 
 In this system  
all the time-odd terms vanish with accuracy higher than the $\stackrel{\leftrightarrow}J^2$  terms, leaving no trace 
on the single-particle spectrum even at the eV level. 
 \\\indent 
 The last remark concerns  the centre-of-mass correction to the kinetic Hamiltonian term.  
 In all our calculations, the total centre-of-mass  of the system at  rest  
 is located at zero - and it is expected to remain fixed if a purely internal excitation is produced. 
\begin{figure}
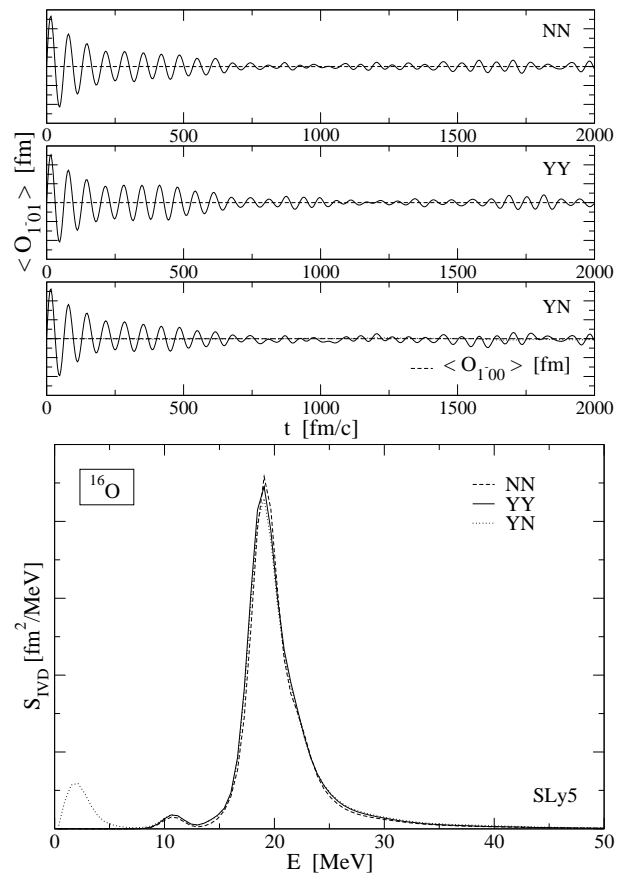
 
\includegraphics[width=8cm]
{Fig1a.eps}
\vspace{1cm} 
\includegraphics[width=8cm]
{Fig1b.eps}
\caption{\label{F1} Isovector dipole (IVD) response in $^{16}$O in the time domain (NN, YY and YN panels) and after the Fourier transform 
 (big  panel). The 
 comparison among calculations including the centre of mass correction everywhere (YY), nowhere (NN) and in the mean-field, but 
 not in the dynamical evolution (YN), is shown. The vertical scale is arbitrary. 
For each calculation, 
  the expectation value of the 
 lowest-order isoscalar dipole operator is reported as well (dashed line).   } 
\end{figure} 
The simple one-body approximation 
 was included 
when 
 fitting the SLy5 force, so one should keep the correction active in the 
static HF. 
 Concerning the dynamics,  it is known~\cite{OU08}-\cite{Erler} that 
 conceptual problems about the definition of the mass of the system 
 arise in TDHF, in relation to 
 processes involving more than one fragment (fusion, fission and collisions in general). 
  When simulating giant resonances in TDHF,  
  one needs to deal with the mass dispersion which accompanies the 
 deexcitation process. 
   Some other authors computing the GDR~\cite{Nakatsukasa} 
  choose to 
  apply the centre-of-mass correction   to the  
   mass distribution in the whole model space, considering 
 the particle number as a 
 constant in the time-space. 
  Other practical recipes can be tried; 
  we  ran 
 a few test cases in order to compare a calculation (YN) including the 
 centre-of-mass   in the static, but not in the dynamic, thus  
 some self-consistency breaking is introduced, a calculation (NN) that  
 neglects it  also in the static   and  the one (YY) 
  accounting 
 for the centre-of-mass  at both stages, with a constant 
 mass number equal to the corrected static one.  
 All these cases  contain  approximations,
 but 
we wanted to outline which is the least worst choice. 
The outcome from this comparison is reported in Fig.~\ref{F1}, where the isovector dipole response in $^{16}$O is 
represented in the time (three horizontal  plots) and energy domain 
(big plot),  
 with  arbitrary vertical scales. 
 The  total centre-of-mass motion (obtained from the expectation value of the 
  lowest-order isoscalar dipole operator 
  $\hat O_{1^{-}00}$, defined analogously to  $\hat O_{1^{-}01}$, but with $\tau_z$ replaced by  
  $I_{\tau}$,   the  identity operator in the isospin space)  
 is plotted as well (dashed line). 
 In the YN calculation 
 the presence of a  
  drift of the system 
 is recorded and it is possible to notice the associated 
   spurious concentration of strength at  low energies.  
 The difference between the 
 NN and the YY calculations 
 turns out to be small with the employed parameters in the linear regime.

\section{Results}
\label{res}
 As previously recalled, the isovector giant dipole resonance,  
   which   characterizes the experimental cross-section   
  more markedly  than other  vibrational states,  
  is  the first nuclear collective excitation which was discovered   
  in photonuclear reactions 
 in the 1940's  (cf.~Ref.~\cite{BM}), 
   imagined by A.~Migdal in 1944~\cite{Mig} in connection to the nuclear matter polarizability.  
  The microscopic interpretation in terms of the
  (overall repulsive)  residual effective interaction, 
 able to explain the collective motion as 
  superposition 
 of nucleons undergoing particle-hole excitations, 
  was achieved only 20 years later.
  Since that time, 
 many experimental data have 
    become available from a wide range of events, 
  together with an extensive  
 theoretical investigation. 
   Forty years after  the first observation,  evidences of
 finite temperature dipole modes  
  taking place in a 
 compound system  from a heavy-ion collision,
  were also found~\cite{New}.  As known, some features of the GDR vary with 
  temperature  (like the size of the deformation and  the 
  width), while the centroid energies tend to remain stable.  
\\\indent 
 The study of the zero temperature modes, the equivalent of the Landau zero sound, 
 represents a  useful way to explore  the nuclear structure 
 and also  to identify features of the intra-medium  
  $N-N$ interaction,  testing how the current models  perform. 
  The main purpose of the current work  is to analyse 
  the effect of the Skyrme-tensor functional on (harmonic)  
   dipole states set on top of the ground-state, 
   although,  
   in the absence of the study of the 
  potential energy dependence on  the deformation,   
  the static mean-field 
  is not guarateed to correspond to the  global  minimum. 
 In particular, in some cases triaxiality is forced for explorative purposes. 
   We leave for the future the investigation of  
    finite  
 temperature  events,  which are important in connection to heavy-nuclei reactions~\cite{Ob2012} 
  and in astrophysical environments. 
\begin{figure} 
\begin{center}
\includegraphics[width=8.0cm]{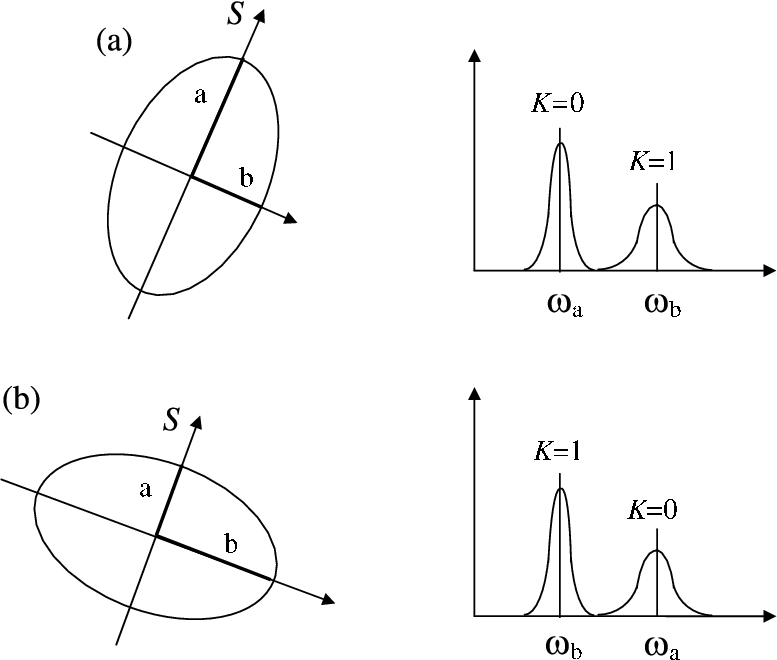}
\caption{\label{F2} Schematic representation of a prolate (a) and an 
 oblate (b) nucleus 
 and the relative  position  of the GDR $K$=0 and $K$=1 centroid energies, with a possible appearance  
 due to the spreading (scales arbitrary; 
 the height of the peak centred around $\omega_b$ needs to be doubled when comparing to the experiment).  
  The half-axis are labelled consistently with  
    Eq.~(\ref{ratio}). 
  The symmetry axis (S) points toward the top. 
}  
\end{center}
 \end{figure} 
\\\indent 
 In this section, the isovector dipole response from TDHF is presented for some 
 benchmark nuclei 
 with a mass number 
  ranging from 16 to 238  
  having, or led to assume, different shapes.   
  As is well known,  the IVD response   in axially    
 deformed nuclei is characterised by a splitting  
  related to the 
  anisotropy of the  
  single-particle   motion:  
  the experimental cross section  
 \begin{equation}
 <\sigma_{1^{-}}>(E)=\frac{1}{3}\sigma_{|1^{-},0\rangle}(E) +\frac{2}{3}\sigma_{|1^{-},  1\rangle}(E) 
 \end{equation}   
is a statistical average  
  of the $|J^{\pi},K\rangle = |1^{-},0\rangle, |1^-, 1\rangle$ modes, 
 respectively   
 associated to oscillations of protons against neutrons
 along ($K$=0) and perpendicularly to ($K$=1) the symmetry 
   axis (see Fig.~\ref{F2}). 
 Analogously, the cross section of giant resonances induced in triaxial nuclei  
  is expected to break up into three contributions  associated to the three 
 characteristic lenghts of the system ($k$).  
 Each distribution    
 can be fitted with a Cauchy-Lorentz 
  probability density function 
 \begin{equation}
 \sigma(E)=\sigma_0 
 \frac{\tilde{\Gamma}^2}{(E-E_0)^2+\tilde{\Gamma}^2},  
\label{Lor} 
\end{equation}
 where $\tilde\Gamma=E\Gamma$, 
 with $\Gamma$ being the  scale parameter, $E_0$ 
  the peak location and $\sigma_0$   the corresponding height. 
   From the  data,    
 one can extract 
 information 
 regarding the type of deformation (triaxial or prolate, oblate shape)  
 and  its size,  respectively represented  by 
  the  so-called Hill-Wheeler parameters 
 $\gamma=\arctan \sqrt{2}\beta_{22}/\beta_{20}$ and 
 $\beta=\sqrt{\beta_{20}^2+2\beta_{22}^2}$, 
   built up from the dimensionless 
 quadrupole deformation coefficients 
$\beta_{2\mu}=\frac{4\pi}{5}\frac{<Q_{2\mu}>}{A<r^2>}$.  
 The splitting between the various $k$ modes can be described by the shift between 
  the centroid energies 
  $\overline{E}_k=m1_{k}/m0_{k}$, 
  computed as   
  the ratio between the energy- and non-energy-weighted sum-rules (EWSR and NEWSR) for each  $k$ component 
 (we will denote the centroid energy of the whole strength distribution by $\overline E$). The splitting     
  is experimentally visible if the nucleus is far enough from the 
 spherical 
 shape ($\beta$=0)  
and 
 if it is not  masked by other effects~\cite{Kolom}.  
  \\\indent 
 A  macroscopic description of the deformation splitting is provided by 
 the hydrodynamical model. 
  Depicting the nucleus as a spheroid with eccentricity $a^2-b^2$, where the semiaxis $a$ 
 is aligned with  the symmetry axis,  
one  has, in particular,  the empirical relation 
for the quantity $\overline E_1/\overline E_0$
(see \cite{BM} and references therein) 
\begin{equation}
\frac{\omega_{b}}{\omega_{a}}=0.911\frac{a}{b}+0.089,  
\label{ratio} 
\end{equation}  
 where $\omega_{b}$ and $\omega_{a}$ are 
 the oscillation frequencies of the liquid drop induced by an external perturbation  
 along the given directions. 
\begin{table}
\caption{Comparison between the  empirical 
   ratios  (second column) 
 of  the dipole $K$=1 and $K$=0 energies, based on Eq.~(\ref{ratio}) and the 
 nuclear size from the static Hartree-Fock approximation, 
 and the corresponding TDHF predictions (third column)  from  the SLy5 
 force, with centroids evaluated in the whole energy range. 
\label{Tab3} }
\begin{ruledtabular}
\begin{tabular}{ccc}
$^{A}$X &  $(\overline E_1/\overline E_0)_{emp}$  & $(\overline E_1/\overline E_0)_{th}$ \\
\hline
$^{24}$Mg & 1.385 & 1.332\\   
$^{28}$Si & 0.757 &  0.758 \\ 
$^{238}$U  & 1.244 & 1.240 \\
\end{tabular}
\end{ruledtabular} 
\end{table}
 Table~\ref{Tab3} 
  compares, for $^{24}$Mg, $^{28}$Si and $^{238}$U, the (SLy5) TDHF centroid energy ratio  with  the right hand side of 
   Eq.~(\ref{ratio}) based on the HF geometry, with the result that 
   the expectation  from this empirical evidence 
  is quite well fulfilled. 
 Similarly,   the centroid energies  can be empirically described in the presence 
  of triaxiality~\cite{GAA}. 
  \begin{figure} 
\vspace{0.3cm}
\includegraphics[width=8cm]{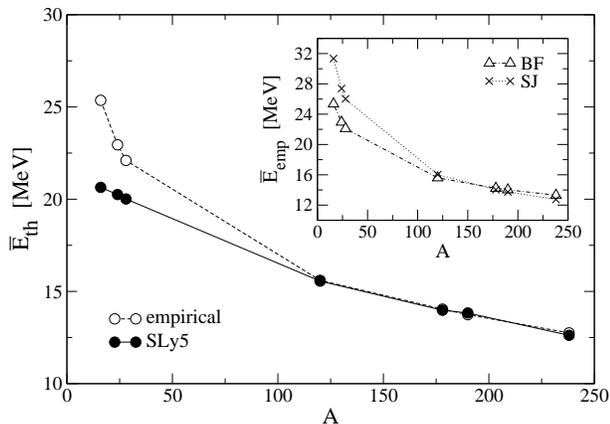}
\caption{\label{F3} Comparison between predicted (full circles) 
 and empirical (white circles) GDR energies through different regions of mass.  
 The theoretical numbers correspond to the total centroids $\overline E$, 
 while  the empirical points are from the BF or SJ model,   as indicated in the main text 
 (the complete empirical  
 trends are shown in the inset).} 
\end{figure} 
\\\noindent Fig.~\ref{F3} shows the 
 comparison between the theoretical centroids (for all the  
 considered nuclei, despite the deformation) and the 
  predictions from empirical fits. 
 The behaviour of the GDR energy when varying  the mass number 
   is    parametrized like 
  $\hbar\omega\sim 31.2 A^{-1/3}+20.6 A^{-1/6}$ according to the Berman-Fultz (BF)  
  model, which is expected to be more appropriate 
 for spherical  light-medium nuclei (typically A$<$100),  where the correction for the surface ($\asymp A^{-1/6}$) is more important, 
 or by $\hbar\omega\sim 79A^{-1/3}$ MeV from  the Steinwedel-Jensen  (SJ) formula (see references in~\cite{Ber},\cite{SJ}).   
  In particular, for each value of the mass number 
  the main plot displays 
  the empirical prediction which is closest to the 
  performance of the SLy5 force (the full trend from both 
  the  models is reported in the inset). 
 The BF formula turns out to be  reproduced better than SJ up to $^{120}$Sn. In any case,  
  it is clear that 
  the discrepancy with the result  from Skyrme   is quite high in the light nuclei;    
     although other sets  can  better perform in such cases,  
 the  simultaneous reproduction of the GDR across different mass regions  
 of the periodic table is a long-standing problem~\cite{Rein}  
   (see also~\cite{Erler} and references therein).  
\\\indent 
As previously anticipated, in this paper  
  various truncations of the SLy5 and T44 Skyrme functional are considered, 
    with the purpose of  studying how the central spin-dependent  and tensor terms manifest  
 themselves   on the nuclear response. 
  Relative effects from the 
 p-h channel are, therefore, discussed. In any case, 
   particular  attention is reserved to the definition  of the pairing force. 
 It should be mentioned that  the predictions for the percentage of total EWSR exhausted in the finite
   energy intervals 
 of interest are quite sensitive to the 
 model parameters. Higher precision calculations are possible in order to provide more robust estimates where needed.  
\\\indent 
Particular  attention is  paid to  the central 
  $\stackrel{\leftrightarrow}{J}^2$ terms  (cf.~Sec.\ref{sk}), 
 which had not yet  
 been included in our model. 
    In this work we will label 
the calculations obtained without the spin-current tensor 
 and with none of the time-odd (with the exception of the time-odd spin-orbit)   
  nor  tensor terms  ``basic'.  
 In all the plots displaying the  
   calculated IVD transition strength  distribution, which is 
  measured in fm$^2$ unless an artificial smoothing is introduced,  
   the scale on the vertical axis has been  normalized  to the highest peak 
  produced, 
 for the  considered nucleus,
  in the   SLy5 
 calculation including the central $\stackrel{\leftrightarrow}{J}^2$  
terms (labelled by J$^2_{\text{c}}$ or simply J$^2$. 
  In the plots where the $K$=0 and $K$=1 components are separately plotted, no relative factor 2 is included). 
 \\
 The results have been extensively verified. 
 Comparisons to other theoretical works or experimental data are provided where available.

\subsection{$^{16}$O}
The first case considered is $^{16}$O, a historically widely used benchmark 
  for  TDHF calculations due to the low demand of 
 computational complexity. 
   Despite 
 the limitations 
 of the mean-field approach and 
 the known  
  problems with 
 reproducing the location of the GDR in light systems,  
 it still is a  good candidate  
to highlight some features  of the Skyrme functional. In fact,  
 the residual interaction 
 is expected to affect 
    how the strength is distributed 
  in light nuclei    more markedly than  in heavier systems, where, by a microscopic point of view,  
 more single-particle levels are involved in the process and 
 shell-dependent effects tend to  average out. 
  \begin{figure} 
\begin{center}
\includegraphics[width=8.6cm]{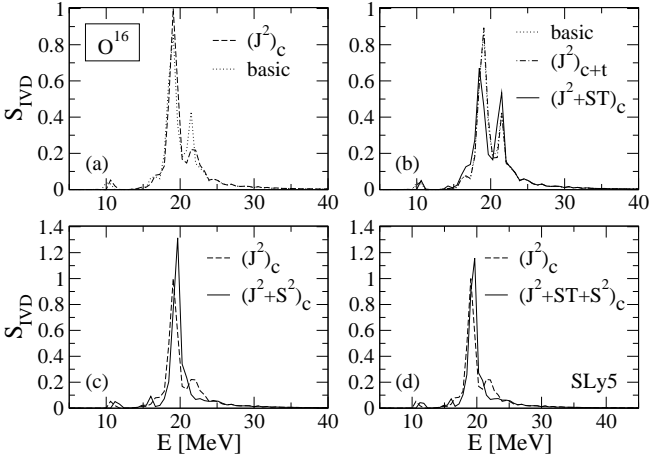}
\caption{\label{F4} Isovector dipole (IVD) strength distribution in $^{16}$O obtained for different truncations of the SLy5 
 functional. The legends indicate the terms made active in the corresponding  calculation, in addition to the ``basic'' version.  
   Details in the main text. }  
\end{center}
\end{figure}
 \begin{figure}
\begin{center}
\includegraphics[width=8.6cm]{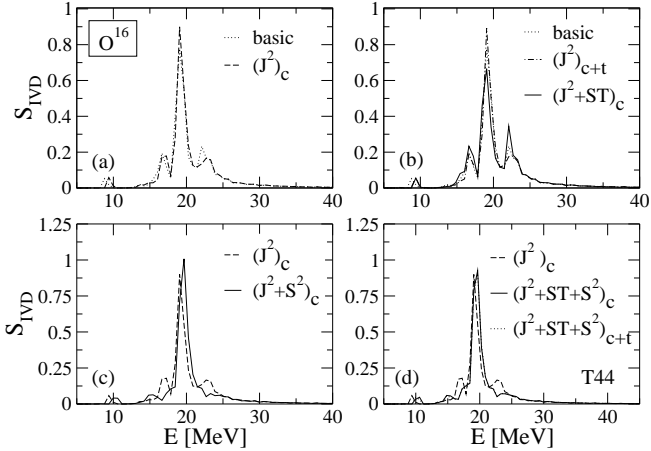}
\caption{\label{F5} Same as Fig.~\ref{F4} for the T44 force.}
\end{center}
\end{figure}
\\\noindent 
The calculated isovector dipole in $^{16}$O  
mainly lies  between 15 and 25~MeV. 
  Panel (a) of Fig.~\ref{F4} shows 
 the response from  the SLy5 functional including  the 
 central  $\stackrel{\leftrightarrow}{J}^2$ 
 terms (dashed line), 
 in comparison to  the ``basic'' calculation previously defined   (dotted line).  
 They 
    produce a  shift to higher energy  of both of the  two main bumps which   
 characterise the response, which are 
   peaked at 
 19.1  and 22.1~MeV.  
 The most prominent effect is obtained  for  the smaller peak at 
 higher energy, the strength of which is reduced by a factor 2 
 in the more complete calculation.  
\\  An evident  change on the strength function appears when including  the 
$\stackrel{\leftrightarrow}J^2$   
Galileian invariant partner 
 in the functional, that is the $\bm S \cdot \bm T$ 
 contribution, which increases the response 
fragmentation for both the Skyrme forces (panel (b)).
  For the SLy5 case, such calculation  produces   two new bumps of almost equal size, peaked 
  at  18.5 and 21.5 MeV. This is the case which qualitatively looks more similar to the 
  experimental cross section, which, in low-resolution experiments,   
  displays two main structures around 22 and 25 MeV~\cite{Ber}. 
  However,   analyses of the microscopic 
   structure  in such 1p-1h limit 
   are  required to add more information. 
  Panel (d) shows  the calculation 
 obtained when including also the $\bm S^2$ terms and confirm their 
  capability to increase the strength of the 
   dipole response,  overcoming 
the effect of the $\bm S \cdot \bm T$  terms.  As a matter of fact, 
 as a consequence of the action of the spin squared terms,   
   there is basically one single centroid   
 with the highest peak at 19.7~MeV,  
    showing  an enhanced transition probability and 
 raised in energy with respect to the J$^2_{\text{c}}$ calculation. 
   The  outcomes described above are similarly obtained  
 with the T44 force (Fig.~\ref{F5}).  
   This ``aggregating'' behaviour of the $\bm S^2$ ingredient was already  observed,  
 without  the $\bm S \cdot \bm T$  terms (c), 
 by Ref.~\cite{Nakatsukasa}, where  the TDHF 
  dipole response in $^{16}$O and  
  Be isotopes  was analysed with the SIII force, but no systematic calculations were carried on. 
  In general, in a p-h view,   
  the spin squared terms (including the  rearrangement due to the density dependence)  
   are  able to couple spin-flip ($\uparrow\downarrow$, $\downarrow\uparrow$), which do 
 not directly contribute to the IVD transition amplitudes,  
 and non-spin-flip  ($\uparrow\uparrow$, $\downarrow\downarrow$) 
  particle-hole pairs with configurations of the same family   
  (and, to some extent, 
    to also  mix them up). 
     By suppressing the $C_T^S$ one-body contribution  that depends on the 
 spin ladder operators, the effect from the spin-squared terms 
 on the whole distribution 
 is lost in our TDHF calculations and 
 one recovers the two-peaked structure of the 
 J$^2_{\text{c}}$ run (the rearrangement has no effect).  
 This finding  is possibly  related 
  to the specific shell structure of $^{16}$O. Further analyses will be discussed in the future.  
  \\ 
 The results obtained by adding  the tensor terms associated to the spin current 
   $\stackrel{\leftrightarrow}J$
      to the  central  ones (hereafter labelled by J$^2_{\text{c+t}}$ like in 
  panels (b)) 
  turn out to be identical,  or displaying a negligible difference,  
 to the ``basic'' version which does not include any of the new terms.  
  The inclusion of the 
  tensor to the $\bm S \cdot \bm T$  contribution  
  does not produce an appreciable effect either. 
\begin{table}
\caption{Isovector dipole centroid energies in $^{16}$O  for various types of calculations
 not including the tensor.    ``Basic'' means that none of the terms specified in the
 subsequent lines have been included. 
 The considered intervals are 
 [17,20] MeV  (I) and (20,24] MeV (II) for SLy5 
([17,22] MeV and  (22,24] MeV for the last two rows),   
 [18,21] MeV for T44 ([17,23] MeV).   
   The percentage of total  EWSR   
 exhausted  in each  energy range is reported within parenthesis. 
  \label{Tab4} }
\begin{ruledtabular} 
\begin{tabular}{cccc} 
 &\multicolumn{1}{c}{SLy5 (I)} & \multicolumn{1}{c}{\text{SLy5 (II)}} & 
\multicolumn{1}{c}{\text{T44}}  \\
\hline 
\text{basic} & 18.94	(52.3\%)&  21.86 (28.2\%)	& 19.40 (52.1\%)\\
$\text{J}^2$ &19.05	(57.0\%)&  22.08 (23.7\%)	& 19.45 (54.4\%)\\
$\text{J}^2\text{+ST}$ &18.71	(43.6\%)&  21.73 (34.8\%)	& 19.34 (45.6\%)\\
$\text{J}^2\text{+S}^2$ &19.66	(75.2\%)& 23.23 (5.20\%)	& 19.72 (72.3\%)\\
$\text{J}^2\text{+ST+S}^2$ &19.56	(74.1\%)& 23.28 (5.90\%)	& 19.62 (71.6\%) 
\end{tabular}
\end{ruledtabular} 
\end{table}
\\The information about the  SLy5 centroid energies for the different types of calculations     
 are summarized in Table~\ref{Tab4},
 together with  the  percentage of total EWSR correspondingly exhausted. 
 They are 
  evaluated in the  intervals 17-20 MeV (18-21 MeV for the T44 force), labelled by (I), where the main peak from the  
 J$^2_{\text{c}}$ calculation is located. 
  The information about the transition strength that is left between 20 and 24 MeV (II) is given as well. 
For the  calculations including the $\bm S^2$ terms  (last two lines), 
the considered interval is 
 17-22 (17-23) MeV: the strength is collected in a single bump absorbing more than the 74$\%$ of the total EWSR. In all the calculations, a small fraction 
 of strength is concentrated around 10 MeV. The total centroid energy amounts to 20.6 MeV (20.8 MeV) for the SLy5 (T44) force. 
\\\indent 
  Studying the  spin response is not the aim of this paper, nevertheless  
   we tracked in time 
the expectation value 
of the  
 operator $O^{K}_{1^-11}=\sum_i Dg_{1K}(\bm x_i,\omega_i)\tau_z(i)$,   
where $g_{1K}(\bm x,\omega)=\mathcal{Y}_{1K}(\bm x)|\bm x| \sigma_z(\omega)$. This corresponds to 
\begin{eqnarray}
  <\hat O^{K}_{1^-11}>(t)=\frac{\tilde D}{2}\int \lambda(S^{(n)}_z(\bm x,t)-S^{(p)}_z(\bm x,t))d\bm x, 
\end{eqnarray}
where $S^{(n)}_z(\bm x,t)-S^{(p)}_z(\bm x,t)$ is 
 the time-dependent spin-isovector density 
$\rho_{11}=(\rho_{n\uparrow}-\rho_{n\downarrow})
-(\rho_{p\uparrow}-\rho_{p\downarrow})$. 
  \begin{figure} 
\includegraphics[width=8.5cm]{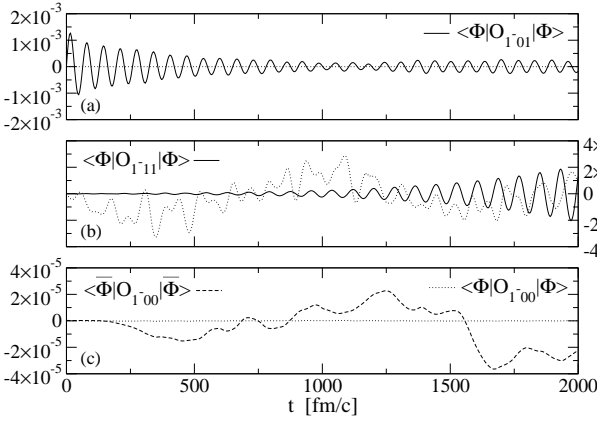}
\caption{\label{F6} Time-dependence of the  dipole response (full line in (a)) and of 
 the  spin-dipole component  (full line in (b)) described  in the main text, for SLy5 in $^{16}$O. 
 The total centre-of-mass displacement (lowest-order isoscalar 
 dipole operator) is 
  plotted as a dotted line in all the three panels. In panel (c), the same quantity corresponding to the 
 worst (YN) case of Fig.~\ref{F1} is reported (dashed line). All the  quantities are directly comparable and measured in fm. }  
\end{figure} 
 The signal takes place at a scale which is several orders of magnitude smaller than the dipole response. 
  For some 
  truncations of the Skyrme EDF, 
  regular oscillations occur from an early stage (Fig~\ref{F6}, full line 
in panel (b)) and appear to grow up in time.  
 The plot is for SLy5 and the same has been found for  other  cases. 
 We  verified that 
      within the time length  considered, and well  beyond,   
 such behaviour, which is not enhanced when reducing the  box size, 
   is  small enough to not influence the observed dipole strength. In particular, the oscillations are not larger than the 
  spurious total centre-of-mass motion, represented, for the same calculation, 
 by a dotted line in all the three panels. This, in turn,  is  two orders of magnitude smaller than 
 the worst case (YN) which was shown in Fig.~\ref{F1} (dashed line in the bottom panel) 
 and comparable to 
   the  NN and YY cases. 
  Further studies are under way.

\subsection{$^{24}$Mg}
The lightest axially deformed system we present is  $^{24}$Mg.  The pairing correlations  
 are negligible and the nucleus has a prolate shape, 
  hence one main splitting of the response arising from the deformation is expected, with the $K$=0 mode, associated to  oscillations along the symmetry 
  (major) axis, 
 being less energetic than the $K$=1 one. 
\begin{figure}
\includegraphics[width=5.5cm, angle=-90]{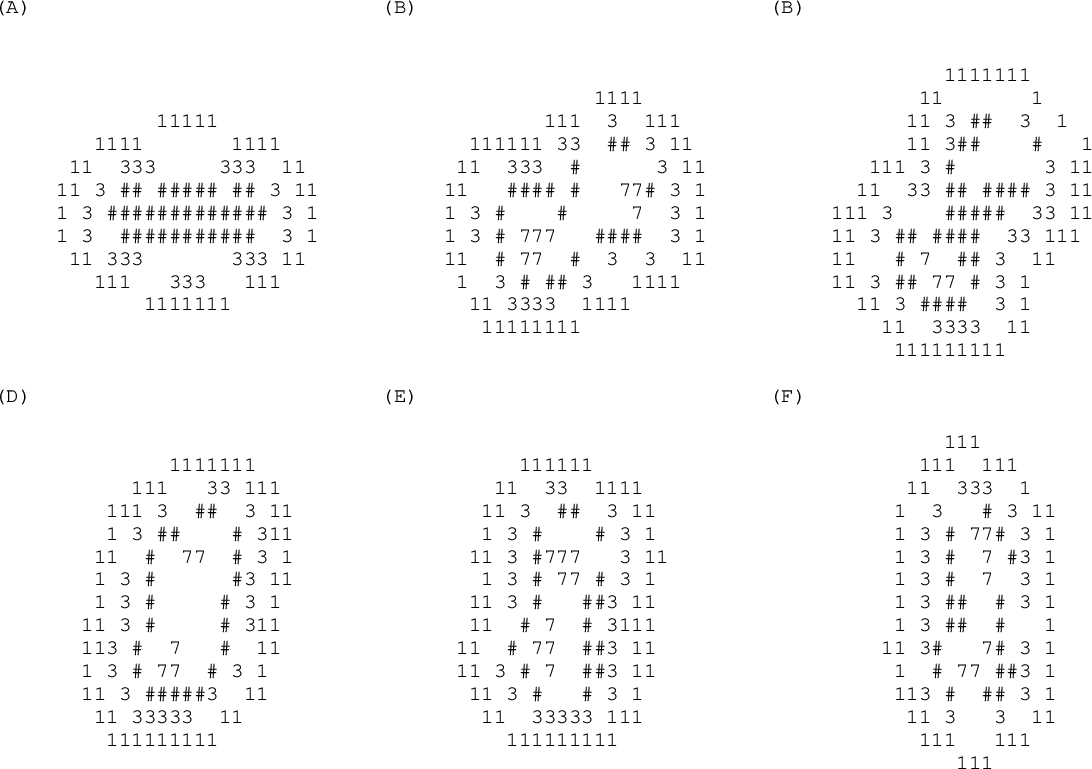} 
\caption{\label{F7c} 
  Density profile in the $x-z$ plane 
 (symmetry axis along the $x$ direction ), describing the 
  late-time evolution (from the top-left to the bottom-right) 
   of the  instability occurring 
 in a J$^2$+S$^2$ calculation in $^{24}$Mg, with the force T44. The 
 cutting plane crosses the middle of the nucleus; the darker 
 label ``\#'' denotes a density of 0.14 particles/fm$^3$, 
 an intermediate value  
 between the contours marked by ``3'' and ``7''.  } 
\end{figure}
 \begin{figure} 
\includegraphics[width=8.5cm]{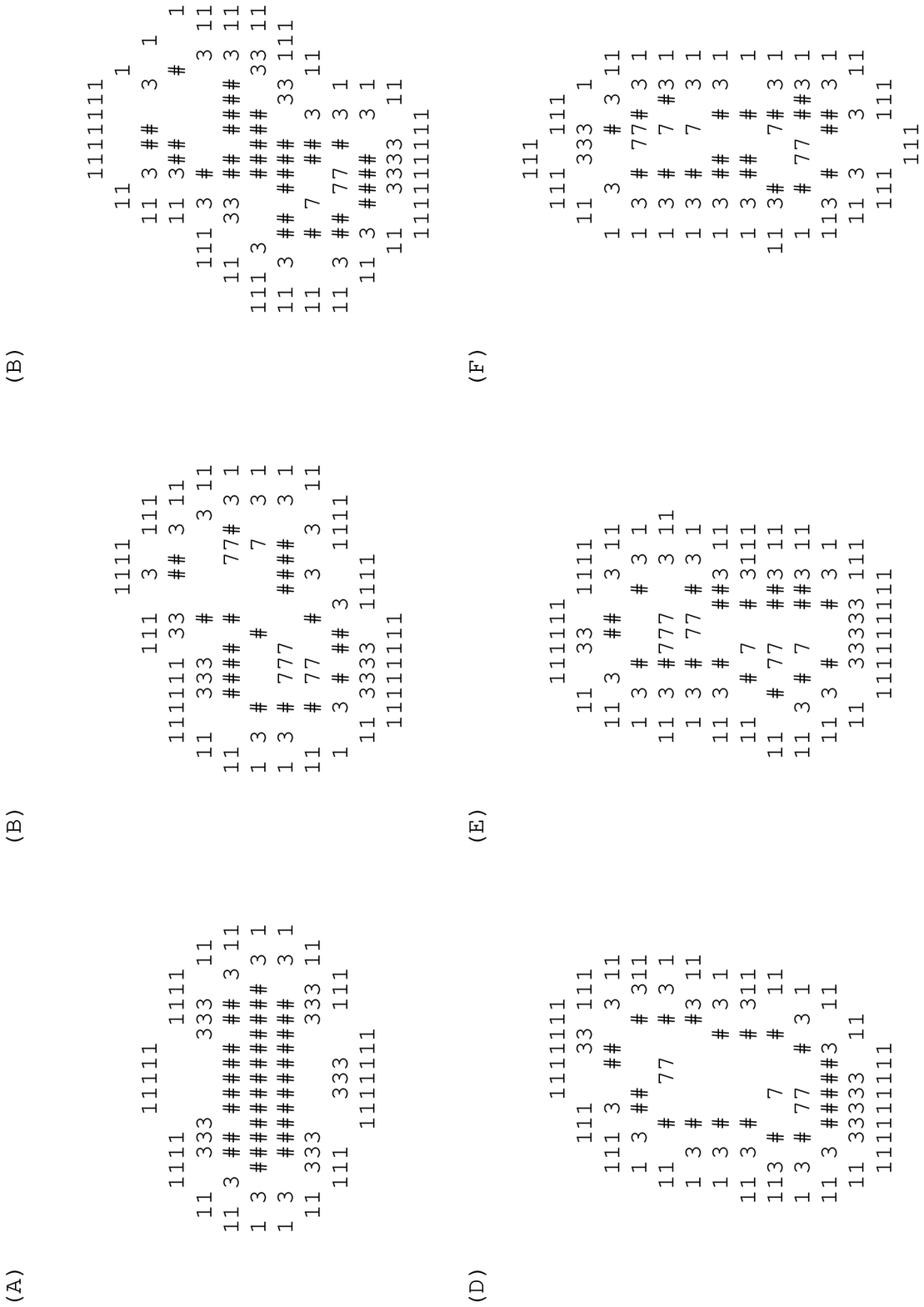}
\caption{\label{F7} IVD response from the SLy5 force in $^{24}$Mg, for 
 the  calculations specified by the labels. Only central terms are included.  
}
\end{figure} 
\\\noindent  The dipole calculations 
 performed with the new terms present some important pathologies 
  when using  the T44 parametrization  and expecially 
 when the boost is applied along the   major  axis: 
  with the new  
 terms, 
  the system displays  spurious excitations, 
   enters the nonlinear regime, and   is led to fission or explosion. 
 This is the case when  adding the tensor from the  
 $\stackrel{\leftrightarrow}J$ spin current: 
 the dipole oscillations are dramatically 
 enhanced and the system splits up 
like as a result of an inelastic collision.   
   Rotational effects can arise as well, as from the 
    J$^2$+S$^2$ model with no tensor (cf.~Fig.~\ref{F7c}), where the dynamics is similar to a fusion state after a noncentral collision.
 With the J$^2$+ST+S$^2$ model, 
    octupole deformations appear on top of the dipole oscillations. 
When the boost is applied along the perpendicular direction,   
   the pathologies   
    due to the inclusion of the $\bm S^2$ terms are largely reduced 
 when the $\bm S\cdot\bm T$ terms are added as well.  
 \\\indent 
  When using the SLy5 force, one would have to   
  run the simulation longer (at least for a time doubled with respect to the T44 J$^2$+S$^2$ case), 
  to face  the same kind of instabilities, which  manifest 
  outside the time window considered in this work, without  affecting 
 the results.  
  These are shown in Fig.~\ref{F7}. 
 Nonzero signal is mainly visible between 13 and 30 MeV, with a narrow peak 
  associated to the $K$=0 mode  located  
at 15.8 MeV (practically unchanged when switching on/off the tensor and time-odd terms of the functional), and a broader bump associated to the second 
characteristic length of the system ($K$=1), 
 with a narrower structure 
 in the 18-23 interval (II). 
   The central J$^2$ model performs similarly to the ``basic'' one. 
For the runs without tensor, 
  the plot presents a pattern similar to Oxygen: 
 the $\bm S^2$ terms 
are 
able to reduce the spreading of the dipole response with respect to the central J$^2$ calculation (dashed line), 
 while  the $\bm S\cdot\bm T$ terms work in the opposite direction (full 
 central calculation J$^2$+ST+S$^2$ represented by a full line).  
    In the inset, the J$^2$+ST calculation is separately drawn.  
  \begin{table} 
\caption{Isovector dipole  centroid energies in $^{24}$Mg  for various types of SLy5 calculations
 not including the tensor,     
  evaluated in the intervals 
 [0,18) MeV (I), [18,23) MeV (II) and in the larger 
 [18,26] MeV window (III). 
 The exhausted percentage of the 
total  EWSR is listed within parenthesis. 
\label{Tab5} }
\begin{ruledtabular}
\begin{tabular}{cccc}
 SLy5 & I  & II   & III \\
\hline 
$\text{J}^2$        & 15.81 (30.0\%) & 20.14 (31.5\%)& 21.78 (56.5\%) \\
$\text{J}^2\text{+S}^2$ & 15.85 (29.9\%) & 20.74 (44.2\%) & 21.42 (56.8\%) \\
$\text J^2\text{+ST+S}^2$ & 15.79 (30.1\%) & 20.58 (42.0\%)& 21.37 (55.5\%)\\
\end{tabular}
\end{ruledtabular}
\end{table}
\\ All the information about   the centroid energies computed in the   intervals of interest   
are listed in 
 Table~\ref{Tab5}, together with 
    the corresponding fraction of exhausted EWSR  
 (see the caption for details). 
 The calculation obtained by adding only the  $\stackrel{\leftrightarrow}{J}^2$ terms to the ``basic'' 
  one places  in the interval II about a 10\% 
less than the  energy-weighted strength of the 
 other cases, because the $K$=1 mode is fragmented at higher energies  
 (still within the III interval).    
  The total centroid  amounts to 20.2 MeV for all the considered 
  truncations of the SLy5 functional. 
\\ The inclusion of the time-odd terms  reduces the overall deformation splitting by 300-400 keV, 
 which globally amounts to about one-half of what is predicted by the Gogny-based QRPA of P\'eru and Goutte~\cite{Peru}.  
  This change is only due to the residual effects in the p-h channel, since  
 the time-odd terms vanish in the ground-state and do not alter the 
 single-particle spectrum.  
  Our outcome is  similar to 
   the SkM*~\cite{Bartel} (Q)RPA result with no tensor presented by Ref.~\cite{Losa} 
(same deformation parameters and r.m.s. are obtained 
 in the ground-state).

\subsection{$^{28}$Si}
  $^{28}$Si is a system comparable to $^{24}$Mg in mass, but with different deformation properties. 
 Both the SLy5 and T44 forces predict an oblate shape ($\gamma$=60$^{\circ}$),  
   with a $\beta$ parameter that, with the second force, is about 30\% weaker when the 
  central  $\stackrel{\leftrightarrow}J^2$ contribution is not included  ($\beta$=0.19 in the ``basic'',  
 versus $\beta$=0.26 of the J$^2_{\text{c}}$ case); it   
 increases 
  by a further $\sim$10\% when  the  tensor is switched on.  The variations  
   are smaller in the SLy5 case ($\beta$=0.28$\pm$0.01 for the three calculations), with  the 
 $\stackrel{\leftrightarrow}J^2$ terms still enhancing the deformation. 
\begin{figure} 
\includegraphics[width=8.6cm]{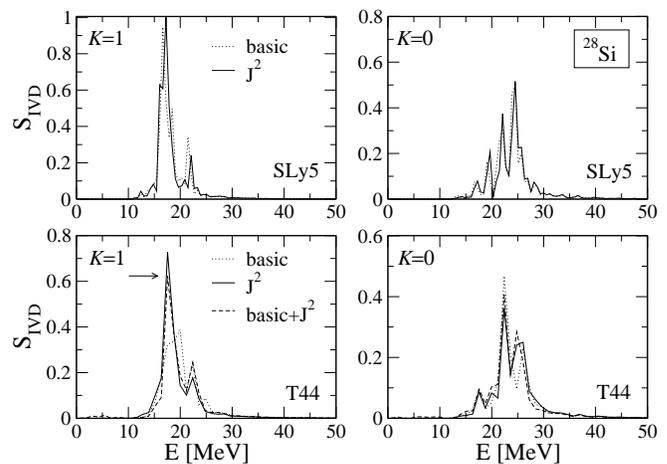}
\caption{\label{F8} $K$=1 (left h.s. panels) and $K$=0  (right h.s. panels) components of the IVD strength 
distribution 
in $^{28}$Si from the SLy5 (top) and T44 (bottom) force. The 
 comparison between the ``basic'' and the (central) J$^2$ calculation is shown. 
 The basic+J$^2$ run has been obtained by suppressing 
 the $\stackrel{\leftrightarrow}J^2$ terms in the ground-state only. 
 The arrow indicates the height of the main peak in the latter case. }
\end{figure} 
 \\  Fig.~\ref{F8} shows how the linear response 
 changes between the   J$^2$ run, where, we recall,  the 
 (central) $\stackrel{\leftrightarrow}J^2$ terms are 
 self-consistently included in both the static and the dynamic, 
  and the  ``basic'' version  where they are neglected at both levels. 
 The upper (bottom) 
 panels are  for the SLy5 (T44) force and 
  the $K$=1 and $K$=0  modes are separately computed 
   at the left- and right-hand side respectively.   
 It is  possible to notice that the response around 22 MeV 
 receives contribution from both the $K$=0 mode and the Landau spreading of the $K$=1 
  component.  
  For the T44 case, the spin-current terms enlarge the splitting  by 
 1.4 MeV, since the $K$=1 centroid lowers from 20.5 to 19.5 MeV, 
 while the $K$=0 one increases by 400 keV, as reported in Tab.~\ref{Tab6}; as a consequence, 
  the ratio between the centroid energies decreases by 6\% (third column). 
  The  empirical estimates of Eq.~(\ref{ratio}) are listed in the last column for comparison. 
  The effect from the 
    SLy5 force is less pronounced, but similar. 
   \begin{table}
\caption{$K$=1  and $K$=0  
 centroid energies from T44 in $^{28}$Si, 
 evaluated in the whole energy range, from the three types of T44 calculations of Fig.~\ref{F8}. 
  The comparison with the  empirical 
  ratio of Eq.~(\ref{ratio}) is given as well.  
 \label{Tab6} } 
\begin{ruledtabular} 
\begin{tabular}{ccccc}
 T44 & $\overline E_1$ [MeV] & $\overline E_0$ [MeV] &  $(\overline E_1/\overline E_0)_{th}$ &  
$(\overline E_1/\overline E_0)_{emp}$   \\ 
\hline
basic & 20.45 & 24.38 & 0.839 & 0.837    \\
J$^2$ & 19.45 & 24.79 & 0.785 & 0.780    \\
basic+J$^2$ & 19.99&23.82 & 0.839 & 0.837 \\
\end{tabular}
\end{ruledtabular}
\end{table}
\\\indent
  As said, a useful tool of investigation of how the energy functional reproduces the nuclear response 
 consists 
       in  performing  
 non-self-consistent calculations 
  where the underlying mean-field is obtained from a functional simpler 
than that used   
 in the dynamic evolution.  In such a way,  
  one can  
 disentangle effects arising from the static 
 mean-field, including the 
      change produced on  the intrinsic deformation. By doing this in the T44 case, it is possible to see that  
  the 1.4 MeV variation of the shift between the $K$=0 and $K$=1 modes 
  can be explained in terms of the 
 change produced on the ground-state, which, as said, 
 is made more oblated by the $\stackrel{\leftrightarrow}J^2$. 
 In fact,  when turning  
    the spin current  on  in a 
 dynamical calculation performed  on top of the ``basic''  mean-field (basic+J$^2$),  
   one obtains a profile similar to the fully self-consistent 
 J$^2$ run (see the dashed lines in Fig.~\ref{F8} for T44),  
  but the gain in the 
 energy splitting is practically lost and one recovers exactly the same energy ratio of the pure ``basic'' case (a small 
 fraction of strength can be noticed at low energies, below 10 MeV, due to the self-consistency breaking). 
   \begin{figure} 
\includegraphics[width=8.6cm]{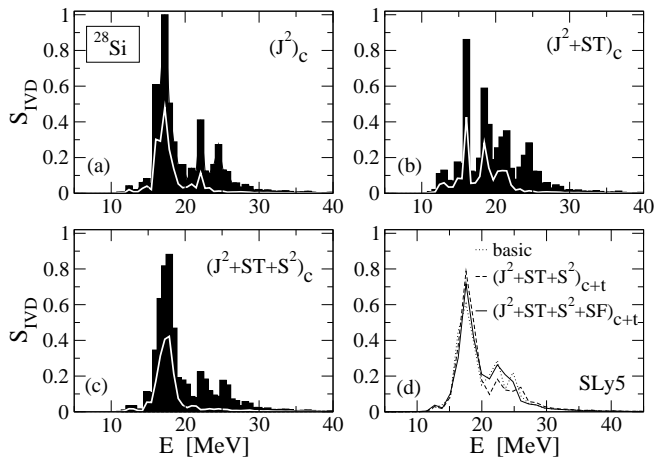}
\caption{\label{F9} IVD response in $^{28}$Si for various types of SLy5 calculations. 
  The white inset  
   shows the (rescaled) profile of the response perpendicularly to the  
 symmetry axis ($K$=1).  
 }
\end{figure}
 \\\indent 
  From  Fig.~\ref{F9}, showing the total IVD strength from the SLy5 force, 
   one can  identify, once again, the opposite behaviour of the 
 central $\bm S\cdot\bm T$ 
 (upper-right  panel) and $\bm S^2$ terms (lower-left  panel)  in 
 redistributing the  transition strength. 
  \begin{table} 
\caption{Isovector dipole centroid energies in $^{28}$Si for the SLy5 force  
 evaluated below (I) and above (II) 20 MeV, for the calculations listed in the first column  (no tensor). 
\label{Tab7} }
\begin{ruledtabular}
\begin{tabular}{ccc} 
SLy5 & I &  II  \\
\hline
 J$^2$ & 17.13 & 25.16\\ 
 J$^2$+ST      & 17.02 & 24.85 \\
 J$^2$+S$^2$ & 17.38     & 25.25     \\
 J$^2$+ST+S$^2$ & 17.40  & 26.02 \\
\end{tabular}
\end{ruledtabular}
\end{table}
  Table~\ref{Tab7} displays the 
  centroid energies computed below (interval I) and above (II) 20 MeV. 
With the $\bm S^2$ ones, strength is pushed at higher energies. 
 The  $\bm S\cdot\bm T$  terms fragment the response, with the result that 
 the main peak is lowered by 100 keV and some transition probability is now visible 
 between 18.0 and 21.5 MeV. 
 \\
 Concerning the tensor, instabilities arise  when including 
the $\bm S\cdot\bm F$ terms. In the case of the SLy5 force, 
 these appear  
 at  the middle of 
  the time 
 length 
 we have typically assumed.  
 We still attempted to compare with 
  runs where the $C_T^F$ coupling constants are  zero by
  halving the simulation time. 
 This operation might lead to loose relevant information, although it turns out that,    
   when repeating the same operation with the cases 
  listed in Tab~\ref{Tab7},  the relative differences 
  between the calculations still hold. 
 From the plot (d) of Fig.~\ref{F9}, 
  it seems  the $\bm S\cdot\bm F$ terms produce an 
  overall attractive  effect.  
 \begin{figure} 
\includegraphics[width=4.7cm]{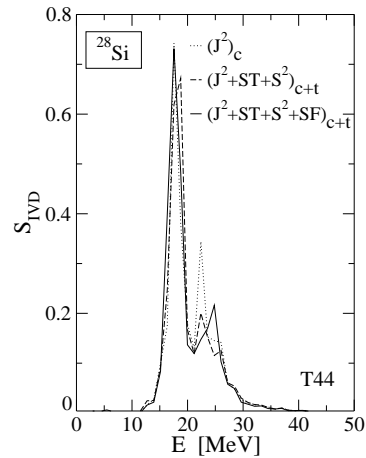} 
\caption{\label{F10} IVD respose in $^{28}$Si for the T44 calculation listed in the legend. }
\end{figure}
The same comparison between T44 calculations  including the tensor is shown in 
   Fig.~\ref{F10}.  
    The $\bm S\cdot\bm F$ terms  shift down the highest peak   
 by a further $\sim$300~keV and also move strength  ($K$=0 mode) to higher energies. 
  In any case, the role of these tensor terms  
 should  be further investigated. 

\subsection{$^{120}$Sn}
The next  case under study  is the spherical $^{120}$Sn, where, at variance with the previous cases, pairing correlations are important and are modelled with the  pairing parameters (set 1) introduced before. 
 Superfluidity 
 opens the channel to the $\stackrel{\leftrightarrow}{J}^2$ 
 tensor terms also in the ground-state, preventing the N=70, Z=50 core from being  
 spin-saturated,   
 although 
 the system remains time-reversal invariant. Consequently, 
  there is  no contribution 
 from the time-odd terms in the mean-field.  
 While checking this,  
 we noticed  that the inclusion of the $\bm S\cdot\bm T$ terms may help 
  the convergence in the static calculation when 
 the squared spin density is turned on.  
 \begin{figure} 
\includegraphics[width=8.5cm]{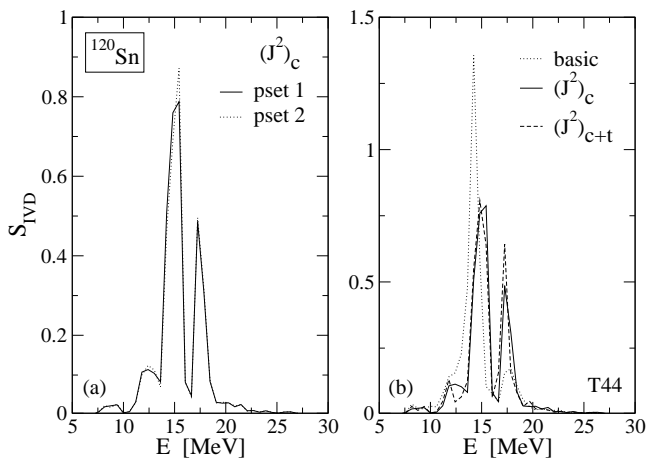}
\caption{\label{F11} Panel (a) shows the dependence of the IVD response in $^{120}$Sn on the pairing set ``1'' and ``2'' defined in the main text, 
 when using the T44 force. Panel (b) shows the effect of the inclusion of the tensor $\stackrel{\leftrightarrow}{J}^2$   
terms (dashed line), in comparison to a calculation including only the central counterpart (full line) and when none 
 of them is taken into account (dotted line).       } 
\end{figure}
 When the $\stackrel{\leftrightarrow}{J}^2$  
 terms are switched on, the pairing gap changes by 1-2\%, which is 
   far from producing  a relevant change on the main  properties of giant resonances. 
 When using the optimum SLy5 pairing set in connection to the T44 force, a pairing gap of 1.84 MeV is obtained. 
 However, even a change of around +20\% of the pairing strength (``pairing set 2''), which  
 significantly alters  the pairing gaps,  
 would produce small variations on the response 
  (cf.~Fig.~\ref{F11}, left panel). 
 There is, consequently, no need to further tune the pairing  
either when modifying the functional for a given force or 
when changing  from SLy5 to T44. 
\begin{figure}
\includegraphics[width=8.5cm]{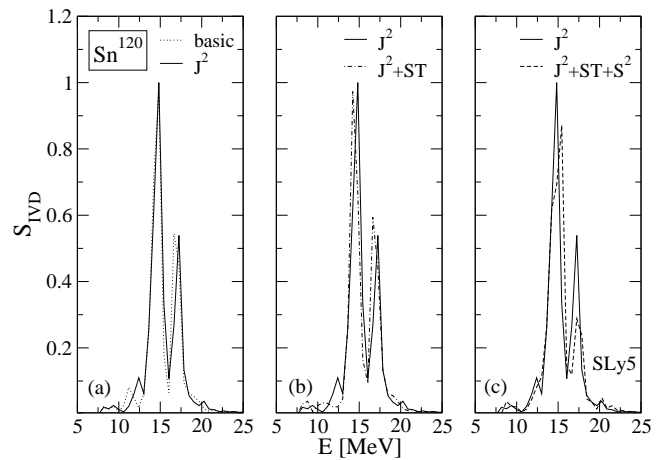}
\caption{\label{F12} IVD response in $^{120}$Sn from the SLy5 force (no tensor); 
 the three panels show the change produced when including the $\stackrel{\leftrightarrow}{J}^2$   
 terms (panel (a), full line),
 when also adding the Galileain invariant partners $\bm S\cdot\bm T$ (panel (b), dashed-dotted line) and when switching on 
 the $\bm S^2$ ones as well (panel (c), dashed line). } 
\end{figure}
\\\indent 
 The SLy5 strength distribution is characterized  by two main peaks 
  (cf.~Fig.~\ref{F12}).  
 The predictions for the  centroid energy ($\sim$15.4 MeV, calculated in the  13.0-18.5 Mev interval)   
   match the 
    experimental value reported in Ref.~\cite{Ber} (cross-section data  fitted 
in the  13-18 MeV range).   
   The 
  change produced by  the new terms in this superfluid A=120 nucleus  is less evident 
  than in the lighter systems. 
  However, one can still notice (b) 
 that the $\bm S\cdot\bm T$ terms, 
    (slightly) reduce  the peaks' height ratio 
 and produce an attractive effect on the centroids, at variance with what obtained   
 when adding  the $\bm S^2$ terms (c). All the calculations modify the low energy tail 
 of the strength function (below 13 MeV), although these outcomes  
    must be considered carefully, 
  since  the low lying states,  
 besides being sensitive to higher order p-h correlations and the pairing,  
    can be particularly   affected by numerical artifacts from the  
  Fourier  transform and spuriousities in the wave-functions. \\
 When the tensor  $\stackrel{\leftrightarrow}{J}^2$   
 and $\bm S\cdot\bm T$ terms are made active, no noticable change 
  on the strength function is introduced when 
 SLy5 is used.   
This is not the case with the T44 force, where the J$^2_{\text{c+t}}$ 
 calculation differs  somewhat 
  (cf.~Fig.~\ref{F11}, right panel).  We finally notice that the two-peak structure always visible with the SLy5 force 
 is lost 
 when using the T44 set in the simplest ``basic'' calculation, where the distribution is shifted to 
 lower energies, but it appears  as soon as 
  the  $\stackrel{\leftrightarrow}{J}^2$    
contribution is added.

\subsection{$^{190}$W}
\begin{figure} 
\includegraphics[width=8.5cm]{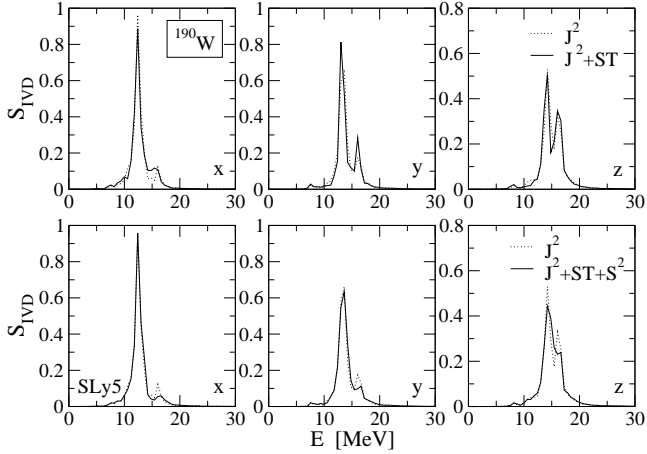}
\caption{\label{F15}
 IVD transition strength  distribution in $^{190}$W 
 on top of a triaxial ground-state. The  $x$ (main axis), $y$ and $z$ components are separately plotted  
 for some truncations of the SLy5 Skyrme functional as described in the main text. }  
\end{figure}

  In triaxial nuclei the deformation  
  leads to a more 
 complex nuclear response when the system is excited.  
 The intrinsic 
 nuclear shape, which impacts on  the 
  dynamical response, 
 results from the balance of 
 different features of the intra-medium $N-N$ force:   
  on one side, various truncations of the Skyrme functional in the p-h channel can affect the deformation properties in the 
 ground-state;  on the other side, (monopole) pairing correlations tend to drive 
 the nuclear system towards a spherical shape and alter the deformation 
 parameters as well,  
  so  
 the proper  focus  on   the pairing 
 force must be posed when studying  intrinsically 
 deformed systems where superfluidity plays a role. \\
  The measurement of the $E_{4^+}/E_{2^+}$ energy ratio, which is 
 a good signature of shape transition,  
  suggests that  $^{190}$W behaves like a triaxial rotor~\cite{ZP}.
When using the 
optimum $^{120}$Sn 
pairing force (set 1) in connection to  the SLy5 functional,  
\begin{figure} 
\includegraphics[width=8.6cm]{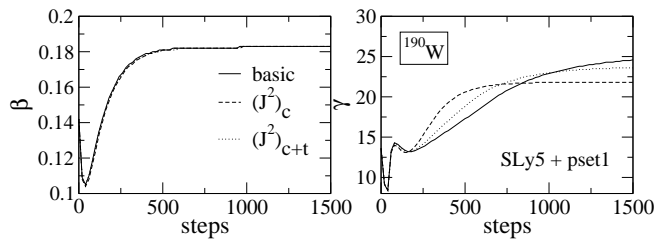}
\caption{\label{F13} Evolution of the deformation parameters $\beta$ 
 and $\gamma$ in $^{190}$W
  during the first stage of the 
 iteration process in Hartree-Fock, for SLy5. Different lines correspond 
 to different runs  complemented by the pairing force 1, as explained in the legend. 
 $\beta$ is adimensional, $\gamma$ is expressed in degrees. 
  } 
\end{figure} 
 a triaxial shape is obtained from our calculations. 
  The two  panels
  in Fig.~\ref{F13} 
  show (an extract of) the evolution  of the 
 $\beta$  and $\gamma$   parameters in $^{190}$W  
 with the number of iterations with the SLy5 functional, 
  for a ``basic'' calculation and when the spin-current pseudo-tensor is included (with and without tensor). 
 This is an example where the 
 inclusion  of the $\stackrel{\leftrightarrow}{J}^2$   terms 
   quicken  the convergence.  
 In particular,  a  14\%  
 decrease  of the $\gamma$  value (21.8$^{\circ}$) with respect to the ``basic'' calculation (25.3$^{\circ}$) is produced. 
  A SLy4 calculation obtained with no  $\stackrel{\leftrightarrow}{J}^2$, with the same starting 
 conditions,   shows a $\gamma$ parameter ($\gamma$=$23.2^{\circ}$) close to   the 
 J$^2_{\text{c+t}}$ SLy5   one ($\gamma$=$23.7^{\circ}$). The $\beta$ parameter does not change.  
   \begin{table} 
 \caption{ \label{Tab8} Isoscalar+isovector  Hartree-Fock energies (in MeV) from the  
   central 
  $J_0^2$, $\bm J^2$ and $\underline{J}^2$ terms of the T44 functional,   
 in the prolate ($\gamma=0^{\circ}$) and triaxial ($\gamma=39.3^{\circ}$) minima  
   found in $^{190}$W. }
\begin{ruledtabular} 
\begin{tabular}{cccc}
 T44   & $\Delta E_{J_0}$  & $\Delta E_{J_1}$ & 
$\Delta E_{J_2}$   \\
\hline
$\gamma=0^{\circ}$ & $\asymp 10^{-7}$ & 11.951 &0.157 \\
$\gamma=39.3^{\circ}$ & $\asymp 10^{-5}$ & 9.671 & 0.105 \\
\end{tabular} 
\end{ruledtabular}
\end{table}
 \\\indent
 The type of truncation of the    energy density functional  
  can not only induce a variation of  the deformation parameters for a given (local or global)
    minimum, but 
  also reverse the relative  location in energy of different minima. 
 For example, 
    when  
    the   $\stackrel{\leftrightarrow}{J}^2$ terms are neglected in 
  the functional based on the T44+pairing 1 parameters,  
   it is possible to identify   one axially deformed  and one triaxial solution, 
   with an energy difference 
   of  0.91 MeV, where  the prolate configuration is  the most bounded.
   By switching those (central) terms on, the situation is reversed, with the  
     triaxial configuration  becoming  
     deeper than the prolate shape   
  (with 
   small variations of the $\beta$ and $\gamma$ 
 parameters, respectively amounting to 
  $\sim$1\% and $\sim$3\%) by 1.42 MeV. 
  This extra binding is mainly due to the fact that the 
 (positive)  energy contribution from 
  the central $\stackrel{\leftrightarrow}{J}^2$ terms    
  in the triaxial case  (9.78 MeV) 
  is 2.3 MeV weaker 
 than in the axial state (cf.~Tab.~\ref{Tab8} where the 
 three 
  $J_0^2$, $\bm J^2$ and $\underline{J}^2$ components are separately shown). \\ 
\\\indent Turning on the time-odd terms, 
  the Kramer's degeneracy is expected to be  preserved. 
  Their  inclusion  
 tends to make the convergence more difficult and 
 the time-reversal invariance is easily broken; in particular,  use of the  T44 force required  effort to find suitable initial conditions. In some cases the problems are emphasised 
 when the $\bm S\cdot \bm T$ terms and the $\bm S^2$ ones are included separately. 
 \begin{figure} 
\includegraphics[width=8.3cm]{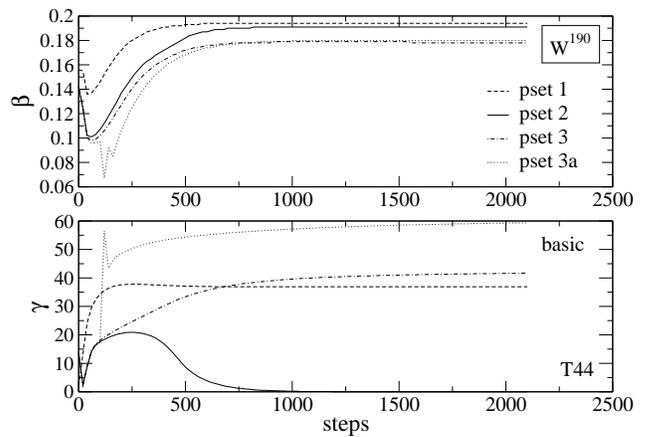}
\caption{\label{F14} Dependence    
 of the deformation parameters  of $^{190}$W 
 on the  pairing force   
 for the T44 Skyrme set with the iteration step number. The 
 pairing sets used in this work are considered. } 
\end{figure}
  We  encountered a  similar situation  when changing the pairing parameters, including 
 the  values of 
 288.5 and 298.8 
 MeV$\cdot$fm$^3$ 
 for the neutron and proton pairing strengths  respectively (``pairing set 2''). 
   These  are optimal parameters in connection to the SLy5 force, 
  as they  provide 
pairing gaps  
$\Delta_{n,p}\in$ 0.7-0.8 
 MeV, 
  much 
 closer 
 to the empirical estimates (0.859 and 0.740 MeV) 
 than the quite small pairing gaps 
 obtained with the pairing set 1 ($\Delta_q\in$~0.2-0.4 MeV, $q=n,p$).  
\\\indent  We notice  interesting 
    variations on the deformation   
  properties when changing the pairing force, with the same initial conditions (the same 
  providing the triaxial solution when using the pairing set 1). 
   Fig.~\ref{F14} shows   the outcome from   ``basic''  T44 calculations  
 based on the  pairing sets 
 considered in this work. 
 If the choice 
 of the pairing set 2   leads, in this case,   to  
 a  prolate axially deformed shape,  
 by reinforcing the proton pairing strength by $\sim$15\%, that is up to 344.00 MeV$\cdot$fm$^3$ (``pairing set 3''), 
 one  triaxial  minimum with $\gamma$=42$^{\circ}$ is constructed.     
  Reversing such  proton and neutron pairing strengths (``pairing set 3a'') 
  would force the system to converge to an 
 oblate shape (dotted line; notice the spike around the 100th step).
   \\\indent 
 Concerning the projection of the linear response in the IVD channel, in a triaxial case 
   we 
 expect a three-modal response associated to the 
 characteristic lengths of the nucleus. 
  Fig.~\ref{F15}  shows the 
  transition strength distribution in $^{190}$W from the SLy5 force,  
  along the x, y and z Cartesian axis and  for some truncations of the Skyrme functional.  
  Each mode 
 appears as a bimodal function, where the highest peak  is respectively 
   located at $\sim$12, 13 and 14  MeV  and a  weaker    
    structure 
 is around 16 MeV in all the panels.  The latter bump     
 becomes increasingly more important from $x$ to $z$ while the main peak reduces in height 
 by $\sim$40\%.    
 Concerning the various functional terms, 
  the $\bm S\cdot\bm T$ ones (three upper panels) 
 produce  an overall  (small)  attraction    
 on the   $y$ and $z$ components.  
 The inclusion of the $\bm S^2$ terms 
  (three bottom panels)
tends, instead, to remove the bimodal structure 
in all the components and the extra mixing places  the (new) centroids  
 at higher energy. \\
\begin{figure} 
\includegraphics[width=8.3cm]{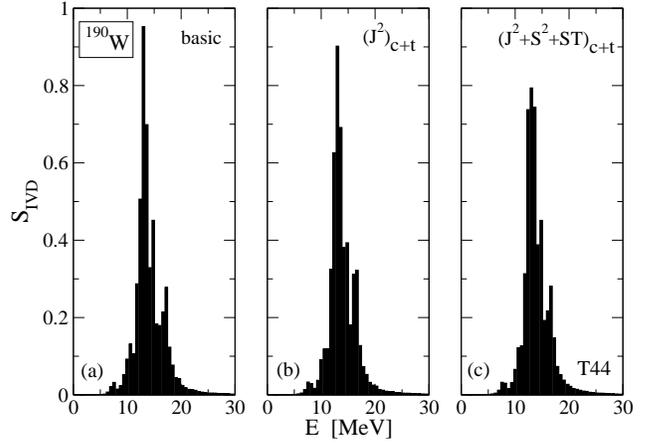}
\caption{\label{F16} 
  IVD strength  distribution in $^{190}$W from the T44 force, in a ``basic'' calculation  and when including the 
 tensor  at the J$^2$  and J$^2$+S$^2$+ST level, as denoted by the legends.}
\end{figure}
 The comparison of the total strength distribution among three types of calculations including the tensor 
  with the T44 force is  shown in Fig.~\ref{F16}. The J$^2_{\text{c+t}}$ calculation (b) and the one including the 
 $\bm S^2$ and  $\bm S\cdot\bm T$ terms as well (c) display a broadening of the response 
 with respect to the ``basic'' (a) case. 
 Although the inclusion of the tensor can help the numerics in some cases, 
 no stable calculation including the tensor 
 $\bm S\cdot\bm F$ terms is currently available in this nucleus.

\subsection{$^{178}$Os}
Another system under study  is $^{178}$Os,  which has been considered a good candidate 
 for shape cohexistance (see \cite{Kumar} and references therein). 
  As for $^{190}$W, 
 we exploit the $\gamma$-softness property to perform  theoretical investigations by forcing triaxiality in some cases.  \\ 
 In this system,  both the empirical proton and neutron pairing gaps belong to the 
 interval  0.91-0.93 MeV. 
   Among the various pairing sets considered, 
 the force 2 still appears the most  suitable, although the convergence is not trivially 
  ensured, expecially for T44.
 \begin{figure}
\includegraphics[width=8.3cm]{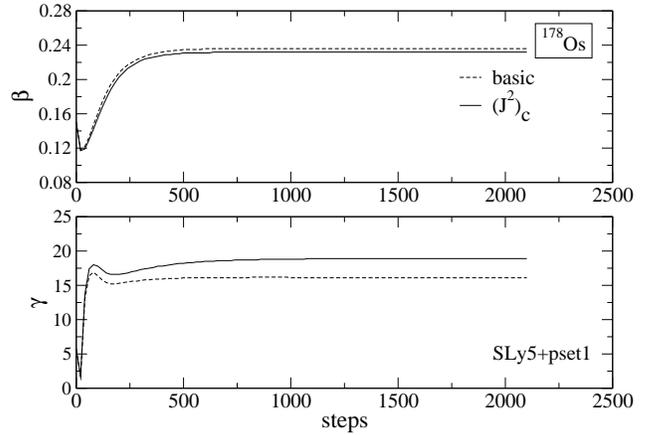}
\caption{\label{F17} Same as
  Fig.~\ref{F13}, for $^{178}$Os.  
 } 
\end{figure}
 The SLy5 force  behaves better also in this system, in terms 
 of convergence.  The HF calculation 
   easily lead to 
   a triaxial minimum;   
  Fig.~\ref{F17} shows 
 that with the pairing force 1 
 (to which pairing gaps $\Delta_n$=0.7 and $\Delta_p$=0.4 MeV correspond),  
 the 
$\gamma$  parameter changes by $\sim$13\%
  when turning the spin-current terms on. 
\begin{figure} 
\includegraphics[width=8.5cm]{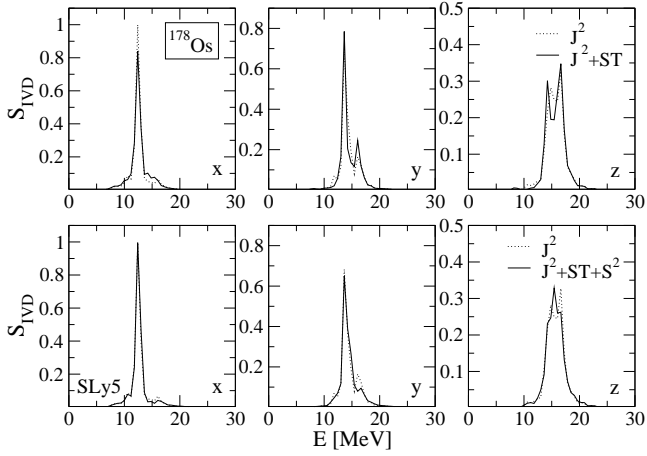}
\caption{\label{F18} Same as Fig.~\ref{F15},  in  $^{178}$Os. }
\end{figure} 
\\\indent 
When performing TDHF calculations on top of the available mean-fields,  
we find outcomes similar to the triaxial case of $^{190}$W 
  (see Fig.~\ref{F18}):   
  the central $\bm S\cdot\bm T$ contribution 
 (three upper panels) 
provides an attractive effect  on the $y$ and $z$ components and 
 increases the spreading
 at variance with   
  the $\bm S^2$ terms (included in the three bottom panels). 
  Along the main axis,  with the $\bm S\cdot\bm T$ terms 
   the signal relaxes down into the tails, which become fatter while the 
    height of the main peak is reduced by $\sim$18\%.    
  \begin{figure}
\includegraphics[width=8.5cm]{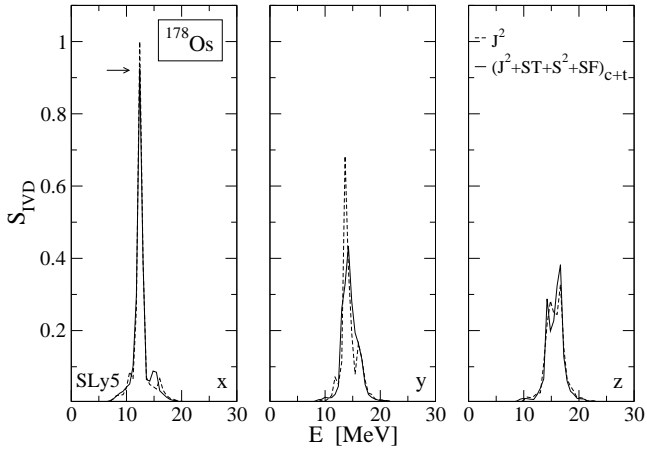}
\caption{\label{F19} Projection of the SLy5 IVD response along the three axis of  $^{178}$Os in the calculations explained in the main text,  
   in comparison to a purely central J$^2$ calculation.  The arrow 
  indicates the height of  the main peak in the former calculation. }
\end{figure} 
 Fig.~\ref{F19} shows the comparison 
 between a purely central J$^2$ calculation and the 
 one including the tensor in a full way (except for  
 the $(\bm \nabla\cdot \bm S)^2$ 
 and $\bm S \cdot\Delta\bm S$ 
terms), 
 with an evident difference in terms of  
   fragmentation, especially on the $y$ component (the total EWSR 
 is still preserved).  
\begin{figure} 
\includegraphics[width=4.6cm,angle=-90]{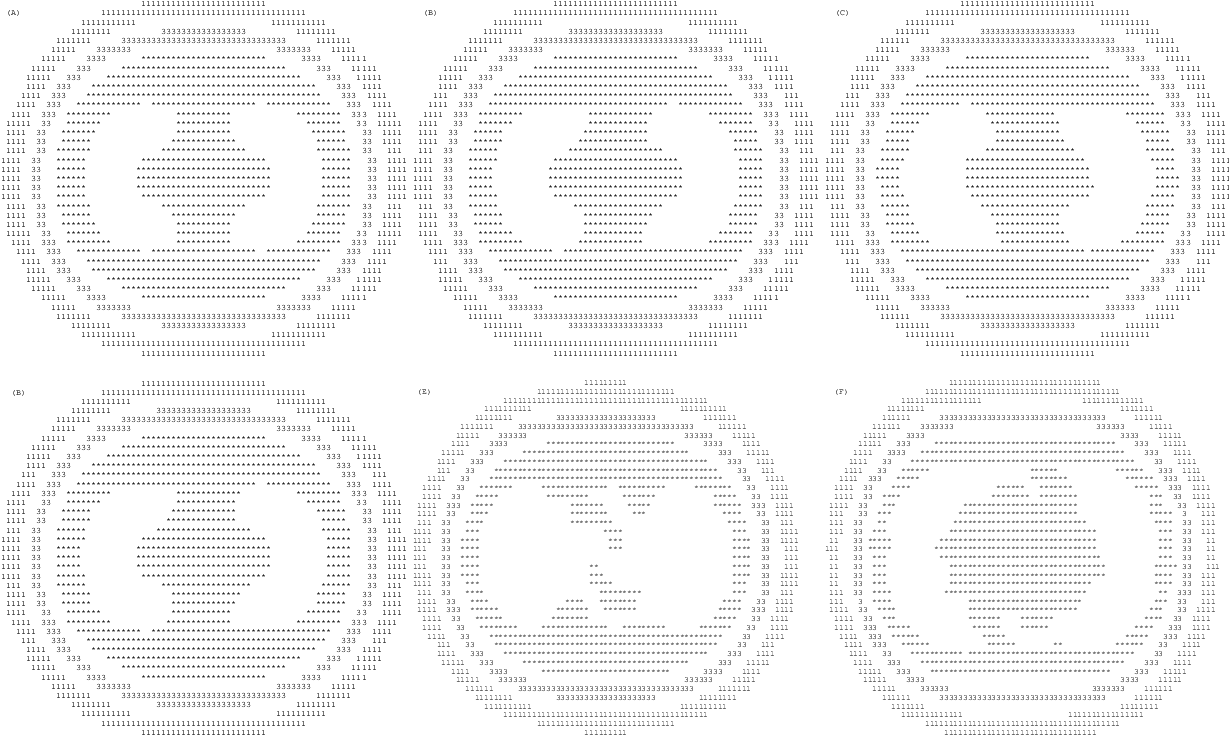}
\caption{\label{F21c} The same as Fig.~\ref{F7c}, 
 for a SLy5 calculation in $^{178}$Os including the central+tensor
J$^2$+S$^2$+ST+SF terms, at a late stage of the dynamics. } 
\end{figure}
  The instabilities associated to long time runs 
 display  rotational currents before the  system expands itself 
 in the whole model space (see Fig.~\ref{F21c}).

\subsection{$^{238}$U} 
The heaviest system  considered in this work is 
the axially deformed $^{238}$U ($\beta$=0.27 with the SLy5 force).  
 \begin{figure}
\includegraphics[width=6.0cm]{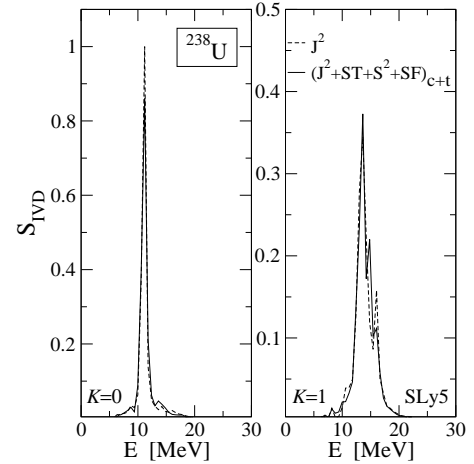}
\caption{\label{F20} IVD strength components in $^{238}$U from the SLy5 force.} 
\end{figure}                  
 The inclusion of the 
 tensor $\bm S\cdot\bm F$
 contribution leads to  a stable result, 
  which is   shown 
 in   Fig.~\ref{F20} (full line) in comparison 
 to a simpler J$^2$ calculation  (dashed line). The  
  $K$=0  and $K$=1 
    modes are separately plotted at the  left and right 
 hand side respectively.   
    \begin{figure}
\vspace{0.2cm}
\includegraphics[width=8.4cm]{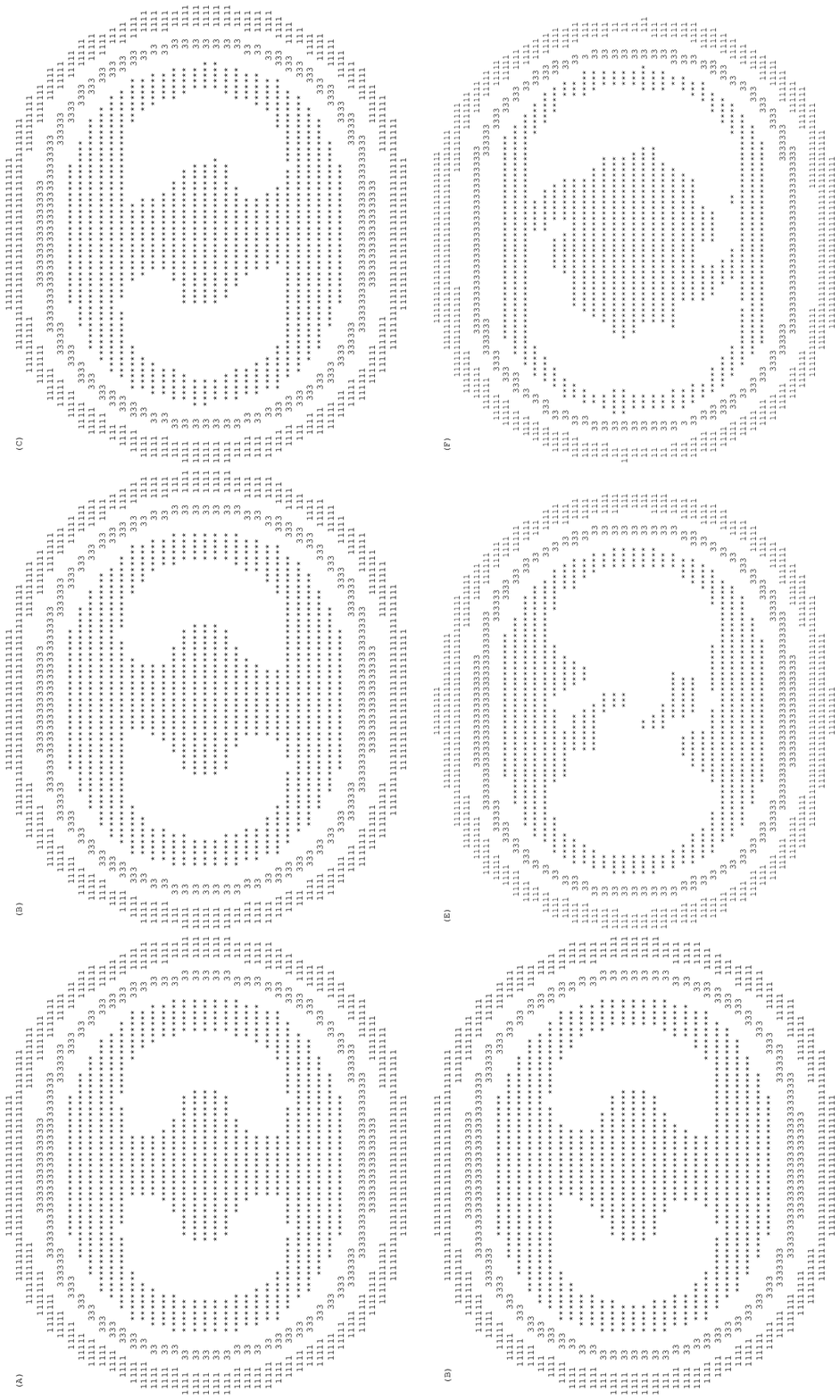}
\caption{\label{F21} 
 IVD response in $^{238}$U from SLy5 with the  tensor contribution.}
\end{figure} 
\\ The resulting smoothed total strength distribution is shown in  Fig.~\ref{F21}. 
  The centroids of the $K$=0 and $K$=1 modes are located at 11.0 
 and 13.5 MeV,  
 not far from the 10.92 and 13.98 MeV reported in Ref.~\cite{Ber}.\\ 
After the smoothing, 
 which simulates the broadening  
 due to higher order effects 
not accounted for in TDHF,   
   the peaks  just above 15 MeV  
 are not resolved any longer.  
 Such level of detail is dependent on the 
 employed box (discretization) and other possible numerical artifacts. 
  However, one cannot   
 exclude that,    when using different parameters, 
 the full treatment of the tensor might more markedly affect 
  the high energy side of the strength 
 distribution, 
 which, in any case,  
  appeared   quite sensitive to the 
    Skyrme set in calculations where the tensor was not included.    
 For example, in Ref.~\cite{Os188}, 
 where a simpler functional based on the SkI3 set was employed, 
 and in 
the  
 separable RPA calculations of Ref.~\cite{def}, based on the   parameters from SLy6,  
  a shoulder in that region  
 was found. 
 Finally, we recall   
 that the modeling of higher order correlations  
    is expected to alter the 
 tails of the 
 strength function. Such extensions  
 with the full Skyrme-tensor functional 
  are undoubtedly 
   envisaged  for the future.

\section{Conclusions}
\label{Concl}
This work presents, for the first time, the full implementation of the 
 Skyrme energy density functional (EDF)  
complemented with the 
  original formulation of the Skyrme-tensor force, 
   in the three-dimensional time-dependent Hartree-Fock (3D-TDHF) 
  with no symmetry restrictions.  
  To our knowledge, 
 this is the most complete  development of TDHF  concerning the p-h channel, 
 where full  
 self-consistency is  achieved between Hartree-Fock and the dynamics (residual interaction) in the 
 sense of no suppression of terms. 
    The derivation of the Skyrme-tensor Hartree-Fock energy  is provided as an extension of Ref.~\cite{Engel}. 
  \\
   The model has been applied in the linear limit, where the random-phase approximation (RPA) is recovered, 
 to simulate the nuclear response 
   in  the  1$^-$ channel, 
 induced 
  by the isovector dipole (IVD) operator. 
  Comments about the methodology have been outlined and some self-consistency issues discussed.  
\\ 
  Various benchmark nuclei 
 with a mass number 
  ranging from 16 to 238  have been considered and triaxial calculations have been presented. 
    The overall results satisfy 
    the empirical trend from the liquid drop model, although, as expected, 
  deviations are found
   in the light region of mass, where the giant dipole resonance (GDR) energy is underestimated. 
  More importantly, 
  the gross features of the  strength distribution from TDHF 
 have been found in agreement with the 
   experimental data  or other theoretical works 
 for all the  systems where information are available.   
\\\indent
     The effects produced by each spin-dependent 
  term of the p-h sector of the functional 
  have been   separately discussed. These are more 
 visible for lighter nuclei 
 where, in the RPA picture,  fewer p-h configurations 
  are involved in the response and Landau fragmentation is more 
 pronounced. 
   However, 
 systematic outcomes are found, for both the  considered Skyrme forces (SLy5 and T44), 
 from $^{16}$O to $^{238}$U: 
   the GDR transition strength, which 
  receives contributions  from one-body transition densities (OBTDs) between parallel spin orbitals,     
     is appreciably sensitive to the central 
 spin-dependent part of the 
  residual interaction.  
   In particular, a    behaviour 
  similar to calculations in other (charge-exchange) channels sensitive to the spin (-isospin) residual force has been noticed.
   This is the case of  the Gamow-Teller resonance, modelled within a RPA formulated in the configuration space~\cite{Fracasso07},  
  the transition strength of which receives contribution from both  spin-flip and non-spin-flip 
  OBTDs. 
  The $\bm S^2$ terms 
 tend to reinforce the residual interaction in the isovector channel, 
 providing a repulsive effect   on average 
 and   enhancing the 
   strength of the resonance  markedly in the considered cases.  
 This dominant 
  mechanism is countered by the  $\bm S\cdot \bm T$ terms,   
  which, although less important in the IVD channel than  what was found in the (charge-exchange) 1$^+$ one, 
  produces  
  more fragmentation in the GR region. 
\\
 On the recalled basis, one would conclude that 
  the  $\bm S\cdot \bm T$ terms, which are spin and  
  velocity-dependent,  
  can   be   useful to balance the  
  $\bm S^2$ contribution of the current  functionals.  
   The former  have been   
    often suppressed in the past, together with their 
 Galilean partners, the time-even  
 $\stackrel{\leftrightarrow}{J}^2$   terms (cf.~the discussion in Ref.~\cite{Fracasso07} and Sec.~\ref{sk}). 
 In  Ref.~\cite{Gor2010},   the removal of the $\bm S\cdot \bm T$ terms was found 
 to improve the performance of 
  the Skyrme EDF in (homogeneous) infinite matter; however, as acknowledged by the same authors,  this suppression  
  would completly remove the $G^{T=0}_1$ and $G^{T=1}_1$  Landau parameters of the standard Skyrme functional,  
  besides spoiling, with more or less manifested effects, the $G^{T=0}_0$, $G^{T=1}_0$  ones, which depend on both the $t_0$, $t_3$ 
   and $t_1$, $t_2$ parameters, respectively through  the $C_T^S$ and 
 $C_T^T$ coupling constants.
  \\\indent 
 Besides influencing the spin-orbit splittings,  the spin-current terms, 
 associated to the pseudo-tensor $\stackrel{\leftrightarrow}{J}$,  
 are of  particular interest 
 in connection to tensor studies,  
  as they allow the tensor to become active even under time-reversal invariant conditions. For this reason, 
   they were 
   the first tensor terms to be considered in the past~\cite{Stancu}, as well as in the more 
 recent history \cite{Les},~\cite{PLB}.\\
  In the ground-state, 
  the inclusion of the  central $\stackrel{\leftrightarrow}J^2$  terms can markedly 
 influence the deformation 
 properties. In particular, 
   variations of the deformation parameters up to $\sim$30-40\% 
 have been recorded in $^{28}$Si, where the splitting between   
 the $K$=0, 1 modes 
 accordingly changes  by $\sim$1.5 MeV, and $^{178}$Os.  
 In $^{190}$W, they  turned out to drive the system to a triaxial shape. 
 The inclusion of the tensor at the   $\stackrel{\leftrightarrow}{J}$ 
  and 
  $\bm S\cdot\bm T$ level does not introduce important changes on the IVD response in the considered cases. 
 A small   effect  is  produced by the $\bm S\cdot \bm F$ terms, 
 but higher precision calculations are required. 
\\\indent 
 Particular 
    attention has been paid to  the pairing correlations, 
  which affect  the intrinsic deformation in terms of both the $\beta$ and $\gamma$ parameters. 
   This has to be taken into consideration when discussing the predictive power of EDF-based calculations. 
   More precisely,  the impact of the smearing of the Fermi surface  produced by 
  a zero range
  (monopole)  pairing force  
  has been discussed: 
  different values of  the pairing parameters have been considered,  
  tuned against experimental masses or exploited as additional degrees of freedom 
 to force the intrinsic nuclear shape.     
 The treatment of the p-p channel does   
  not alter our conclusions about the p-h channel, where relative 
 effects are mainly considered, and do not prevent the comparison with the experiment. 
\\\indent 
 The occurrence of  instabilities from the Skyrme functional 
    has attracted  particular interest in recent times (cf.~\cite{Kortelainen} and 
 references therein). 
 In our calculations,  various types of instabilities 
 from the spin-density dependent terms of 
   the SLy5 or T44 functional  
  have been found, which are 
  enhanced in the presence of derivatives, depend 
  on  the way the  EDF terms couple one another  and   mix with 
    the  zero  modes. 
 They appear similarly to the possible cases described in Ref.~\cite{Kortelainen}.  
 This is especially true when the T44 force is employed, while the SLy5 set  behaves more regularly.
    A peculiar  case  is represented by the performance of T44 in $^{24}$Mg, 
 where  different truncations of the functional 
 allow the spuriousities  to dominate the scene in a relatively short time.  
   This might be related to possible softness properties of this nucleus, which 
 are predicted by some calculations~\cite{Losa}. 
   The T44 force, moreover, 
   turned  out   to    hinder  the convergence in the static Hartree-Fock quite easily in our calculations,  
  making the numerics more difficult. \\ 
  We notice that 
   the $\bm S^2$ and the $\bm S\cdot \bm T$ terms  can produce an unphysical time-reversal breaking 
  in both the ground-state and the long term dynamics of the heavy $\gamma$-soft nuclei we considered. 
 The inclusion of both those EDF contributions turned out to reduce the appearance of unwanted effects in some cases, with respect 
 to the situation where only one of them is included. 
  It similarly happens when  the tensor is added to the central $\bm S\cdot \bm T$ terms,  
  due to cancellations. 
  The terms depending on the 
 laplacian   and on the divergence of the spin density  
   have not been analysed in this work. 
 No instabilities  related to the laplacian of the particle density or to the other 
  time-even contributions  have risen up, but more  focused and systematic analyses in both  sectors 
  can be accomplished. 
\\
Unphysical dissipation effects in TDHF,  
 when time-odd terms from the spin-orbit sector of the functional 
 are suppressed, were observed in Ref.~\cite{PRE}  and 
  divergences taking  place when running the 
  simulation for a long time, in   relation to  couplings with zero-modes,   
 were mentioned in Ref.~\cite{Nakatsukasa}. 
   A more detailed study of such phenomena 
  is envisaged     
 also in view of computing 
   the spin modes and we will deal with this subject 
 in a separate work.  
 The 3D-TDHF 
  could  represent a   particularly sensitive testing tool for 
   Skyrme(-like) energy density functionals in finite systems and complement information from infinite matter 
 analyses~\cite{Davesne}-\cite{Pastore}. 
 Whereas the problem of instabilities might represent a limitation  
 and demands for a solution that does not  
   oblige to drop 
   terms  by possibly introducing further spuriousities, besides causing 
 possible lack of physical effects, 
 useful information 
 about the nuclear  structure  can be accessed  
 through  correlations associated to the symmetries breaking.  \newline
\indent
Changing the functional by dropping terms according to one's needs is  
 useful for explorative purposes,  
  the ultimate goal of this line of research still  being 
 the realization 
 of a reliable 
 functional, ideally flexible enough 
 to adapt to disparate conditions.   
   Finding the proper balance between the various spin-dependent terms 
   would improve the predictive power of the current formulations.     
   As is known,   
  one conclusion of  
 Ref.~\cite{Les}  was that   the available parametrizations for the tensor force are
   not able to adjust the drawbacks of the central and  spin-orbit part 
 of the Skyrme functional, although it can offer extra degrees of 
 freedom to fix the velocity-dependent terms.  
    Interest is still driving the search for suitable 
   constraints sensitive to the spin-dependent terms 
  from  excited states and/or ground-state properties. 
    Clearly, in order to ensure predictability on collective excitations in the RPA or equivalent approaches  
  that strongly depend on the underlying shell-structure, 
 coherent     efforts are required  to improve  
     the description of the involved single-particle degrees of freedom. 
 \\ 
 The GDR energy is known to be correlated to the symmetry energy  
  in dependence on the TRK enhancement factor, expressed in terms of 
   the $t_1$ and $t_2$ 
 parameters through the effective mass~\cite{Trippa}, which is already 
 employed in fitting procedures (see, e.g.,~Ref.~\cite{Chabanat}).  
    Input from the deformation splitting could  also be  taken into account.    
\\\indent 
Concerning further applications,
 we saw  that the spin-dependent terms of the Skyrme functional  
 can affect the tails of the nuclear response. 
The fraction of dipole strength concentrated in the low energy tail of the GDR, around the 
  threshold for particle emission, impacts on the photodisintegration rates,  
    of particular interest  for  nucleosynthesis (see, e.g.,~Ref.~\cite{Utsu}). 
   In such region, also pertaining to pigmy resonances,   pairing correlations and 
    the coupling to low-lying phonon can  play a role~\cite{DS}.
   The investigation of such an energy window  by means of TDHF simulations, even in the linear approximation,  
    is delicate, because proper attention is required to remove artifacts related to the Fourier transform 
 and possible unphysical effects from the coupling to zero modes. 
 In any case, 
 the treatment of anharmonicities with the full Skyrme-tensor functional, in particular the extensions 
 to account for coupling to 
 high lying 2p-2h states, which is expected to be strengthened by the tensor force, would be an interesting step 
 to  undertake. \\\indent
Finally, 
 studying  
 particle emission  from  collective excitations 
     allows one to access information about the involved single-particle degrees of freedom  
  as well as the mechanisms responsible for their cooperation,  
  of interest 
 for both the (interdependent) 
 nuclear structure and  nuclear interaction modeling.

\begin{acknowledgments}
 The authors would like to acknowledge 
 Prof.~Ron Johnson, 
 Dr~Alexis Diaz-Torres and Prof.~Nikolay Minkov 
 for useful discussions and suggestions at an early stage of this work.  
  Prof.~Joachim A.~Maruhn is also acknowledged for reading of the manuscript. 
 \end{acknowledgments}

\appendix
\begin{widetext}
\section{The unrestricted Skyrme-tensor Hartree-Fock energy density functional} 
\label{App1}
 In this Appendix, the 
  derivation of the Skyrme  energy density functional provided in Ref.~\cite{Engel}     is extended 
 in order to include the tensor contribution,   
 under the general hypothesis of no symmetry restrictions. The expression 
 of the zero-range   tensor force as formulated in Ref.~\cite{Stancu} is employed 
 (cf.~Ref.~\cite{PRC2012} for the equivalent derivation of the functional based on Eq.~(\ref{vtens}), leading to the 
 same result). 
 For the sake of  convenience, we re-write  it here as 
 \begin{eqnarray}
v_{\tau}(1,2) 
&=&\nonumber\\ 
&&\hspace{-0.5cm}\left\{\frac{T}{2}\Big[(\bm{\sigma}_1 \cdot \bm{k}')(\bm{\sigma}_2 \cdot \bm{k}')\delta(\bm{x}_1-\bm{x}_2)+
\delta(\bm{x}_1-\bm{x}_2)(\bm{\sigma}_1\cdot \bm{k})(\bm{\sigma}_2\cdot \bm{k})\Big]-\frac{T}{6}(\bm{\sigma}_1\cdot\bm{\sigma}_2)\Big[\bm{k}'^2\delta(\bm{x}_1-\bm{x}_2)+\delta(\bm{x}_1-\bm{x}_2)\bm{k}^2 \Big]+\right.\nonumber\\
&&\hspace{-0.3cm}
\left.+U(\bm{\sigma}_1\cdot\bm{k}')\delta(\bm{x}_1-\bm{x}_2)(\bm{\sigma}_2\cdot\bm{k})\
-\frac{U}{3}(\bm{\sigma}_1\cdot\bm{\sigma}_2)\Big[\bm{k}'\cdot\delta(\bm{x}_1-\bm{x}_2)\bm{k}\Big]\right\}(1-P_{\sigma}P_{\tau}P_{M}), 
\label{eqn}
\end{eqnarray}
where 
$\bm k=\frac{1}{2i}(\bm\nabla_1 -\bm\nabla_2 )$ and $\bm k'=-\frac{1}{2i}
(\bm\nabla'_1- \bm\nabla'_2)$, respectively, act on the right and on the left, 
$\bm \sigma$ are the spin Pauli matrices 
and $P_{\sigma}, P_{\tau}, P{_M}$ are the usual operators that exchange the spin, isospin and spatial coordinates. 
 When computing the Hartree-Fock energy~(\ref{EHF}),   the $(1-P_{\sigma}P_{\tau}P_{M})$ operator allows, like for the central and spin-orbit terms, 
 to account for the exchange without antisymmetrizing the 
  wave-function in the ket. 
 For the considered force, and under the hypothesis of no charge mixing, 
 the  product of the three exchange operators can be replaced by $\pm \delta_{ q_1, q_2}$ 
 ($q=p,n$ )
 for the $T-$ and $U-$ weighted terms, respectively. 
 To emphasize the spatial behaviour  
of the corresponding interaction terms, 
 the tensor parameters are defined  
 as $T=3t_e$ and $U=3t_o$ in some works. 
\\\indent 
Although we deal with a zero range force, 
 it is appropriate to render the dependence on space 
  more explicit  in order to proceed with the 
 calculation.  
By representing the  $\sigma_z$  
 diagonal 
  basis 
 by the set $\{|\omega\rangle\}$, where $\omega=2m_s=\pm 1$ identifies the  $\uparrow,\downarrow$ eigenvectors, 
  the direct contribution to the Hartree-Fock energy from the first two terms of Eq.~(\ref{eqn}) 
is computed  as 
\begin{eqnarray}
\bigtriangleup 
E^D_I&=&-\frac{T}{16}
\sum_{i,j}\sum_{\omega_{1'},\omega_1,\omega_{2'}\omega_2}
\int d\bm R\Big\{  
 \phi^*_{i}(\bm x'_1,\omega'_1)\phi^*_{j}(\bm x'_2,\omega'_2)\delta(\bm x_1-\bm x_2) 
\delta(\bm x_1-\bm x'_1) \delta(\bm x_2-\bm x'_2) 
\nonumber \\
&&\Big[
\langle \omega'_1 | \bm \sigma_1 | \omega_1 \rangle
\cdot(\bm \nabla_1-\bm\nabla_2)\Big]
\Big[  
\langle \omega'_2|\bm \sigma_2|\omega_2\rangle\cdot
(\bm\nabla_1-\bm\nabla_2)\Big]
\phi_i(\bm x_1,\omega_1)\phi_j(\bm x_2,\omega_2)\Big\}+\mbox{H.c.},  
\end{eqnarray} 
where H.c. denotes the Hermitian conjugate of what preceeds and $d\bm R$ stands for  $d\bm x'_1d\bm x'_2 d\bm x_1 d\bm x_2$ 
(in general, $\delta(\bm x_1-\bm x'_1) \delta(\bm x_2-\bm x'_2)$ ensures the force has a local character).  
 The subscripts i, j   
 summarize the quantum numbers 
of the spinor  $\psi_i(\bm x)=\sum_{\omega=\pm} 
\phi_i(\bm x,\omega)|\omega\rangle$; 
 the isospin is represented through the charge $q$, among the other quantum numbers. 
In the following, unless differently specified, all the 
  derivative operations 
 apply to 
  the right. \\
By inserting the definition 
 of the spin-density~(\ref{dens2})  
$\bm S_q(\bm x)= \left.\bm S_q(\bm x,\bm x')\right.|_{\bm x=\bm x'}$, 
where, like for the other densities,  
 $\bm S(\bm x)=\sum_{q=p,n} \bm S_q(\bm x)$
denotes the 
  isoscalar   
   density  $\bm S_{T=0}(\bm x)$,   
 the previous energy contribution can be recast as
\begin{eqnarray}
\bigtriangleup E^D_I&=&-\frac{T}{16}\sum_{\mu,\nu}
\int d\bm R\Big\{\Big[
 2S_{\nu}(\bm x_2,\bm x'_2)
\nabla_{1,\mu}
\nabla_{1,\nu}
S_{\mu}(\bm x_1,\bm x'_1)-
\nabla_{1,\mu}S_{\mu}(\bm x_1,\bm x'_1)
\nabla_{2,\nu}
S_{\nu}(\bm x_2,\bm x'_2)+\nonumber\\
&& -
\nabla_{1,\nu}S_{\mu}(\bm x_1,\bm x'_1)
\nabla_{2,\mu}
S_{\nu}(\bm x_2,\bm x'_2)\Big]  
+\mbox{H.c.}\Big\}_{\bm x_1=\bm x'_1=\bm x_2=\bm x'_2}\hspace{-1.9cm}.
\end{eqnarray}
By inserting the relation~(A.7) of~\cite{Engel}
\begin{equation}
\left[
{\nabla}_{\mu}S^{(q)}_{\nu}(\bm x,\bm x ')
\right]_{\bm x=\bm x'}
=
\left[
{\nabla'}_{\mu}S^{(q)}_{\nu}(\bm x,\bm x ')
\right]^*_{\bm x=\bm x'}
=
\frac{1}{2}
{\nabla}_{\mu}S^{(q)}_{\nu}(\bm x)+iJ^{(q)}_{\mu\nu}(\bm x)   
\label{1E}
\end{equation}
   and the identity  
\begin{equation}
\sum_{\mu\nu}
S^{(q)}_{\nu}(\bm x_2,\bm x'_2)
\left(
\nabla_{1',\mu}
\nabla_{1',\nu}+
\nabla_{1,\mu}
\nabla_{1,\nu}\right)
S^{(q)}_{\mu}(\bm x_1,\bm x'_1)
=2
 \bm S_q(\bm x_2,\bm x'_2)\cdot 
\bm G_q(\bm x_1,\bm x'_1),  
\end{equation}
 where 
  $\bm G(\bm x)=\bm G(\bm x,\bm x')|_{\bm x=\bm x'}$ has been firstly introduced in Eq.~(\ref{GG}),  
  the energy becomes 
\begin{eqnarray}
\bigtriangleup  E^D_I&=&-\frac{T}{8}\int d\bm x_1 d\bm x_2 
\Big
\{2\bm S(\bm x_1)\cdot\bm G(\bm x_2)-\frac{1}{4}\Big[\bm \nabla_1\cdot \bm S(\bm x_1)\Big]
\Big[\bm \nabla_2 \cdot\bm S(\bm x_2)\Big]  
+J_0(\bm x_1)J_0(\bm x_2)+\nonumber\\
&&-\frac{1}{4}\sum_{\mu\nu}
\nabla_{1,\mu}  S_{\nu}(\bm x_1)\nabla_{2,\nu} S_{\mu}(\bm x_2)
+\sum_{\mu\nu}
J_{\mu\nu}(\bm x_1)J_{\nu\mu}(\bm x_2)\Big\}_{\bm x_1=\bm x_2}\hspace{-0.8cm}. 
\end{eqnarray}
  Performing integration by parts twice on the fourth term, after also the $\delta(\bm x_1-\bm x_2)$ function has acted, 
and using, in each point of space,  the identity 
\begin{equation}
\sum_{\mu\nu}J^{(q)}_{\mu\nu}J^{(q)}_{\nu\mu}
=
\underline J_q^2 -\frac{1}{2}\bm J_q^2+\frac{1}{3}(J^{(0)}_q)^2   
 \label{eqJ_}  
\end{equation} 
for the last term, leads to 
\begin{eqnarray}
\bigtriangleup  E^D_I&=&T\int d\bm x 
\Big
[-\frac{1}{4}\bm S(\bm x)\cdot \bm G(\bm x)+\frac{1}{16}(\bm \nabla \cdot\bm S(\bm x))^2-
\frac{1}{6}{J_0}^2(\bm x)-\frac{1}{8}\underline{J}^2(\bm x)+\frac{1}{16}\bm J^2(\bm x)\Big]. 
\label{first}
\end{eqnarray}
By similarly  proceeding  
for the remaining two terms of the tensor force that are weighted by the parameter $T$, one gets 
\begin{eqnarray}
\hspace{-0.7cm}\bigtriangleup  E^D_{II}&=&\frac{T}{24}\sum_{\mu,\nu} 
\int d\bm R 
\Big\{\Big[S_{\nu}(\bm x_2,\bm x'_2)
(\nabla_{1,\mu})^2
S_{\nu}(\bm x_1,\bm x'_1)-
\nabla_{2,\mu}
S_{\nu}(\bm x_2,\bm x'_2)
\nabla_{1,\mu}S_{\nu}(\bm x_1,\bm x'_1)\Big]
+\mbox{H.c.}\Big\}_{\bm x_1=\bm x'_1=\bm x_2=\bm x'_2}\hspace{-1.9cm}.  
\nonumber\end{eqnarray} 
By using the relation~(A.6) of~\cite{Engel} 
\begin{eqnarray}
\Big[(
\bm\nabla^2+
\bm\nabla'^2)S^{(q)}_{\mu}(\bm x,\bm x')\Big]_{\bm x=\bm x'} &=&
\Delta S^{(q)}_{\mu}(\bm x)-2T^{(q)}_{\mu}(\bm x),
\end{eqnarray}
 in addition to  Eq.~(\ref{1E}), one  obtains 
\begin{eqnarray}
\hspace{-0.4cm}\bigtriangleup  E^D_{II}&=&\frac{T}{24}\int d\bm x_1d\bm x_2 
\Big\{\bm S(\bm x_2)\cdot\Delta \bm S(\bm x_1)-
2\bm S(\bm x_2)\cdot\bm T(\bm x_1)
-\frac{1}{2}\nabla_{2,\mu} S_{\nu}(\bm x_2)\nabla_{1,\mu} S_{\nu}(\bm x_1)
+2\sum_{\mu\nu}J_{\mu\nu}(\bm x_2)J_{\mu\nu}(\bm x_1)\Big\}_{\bm x_1=\bm x_2}\hspace{-0.8cm}.    
\nonumber
\end{eqnarray}
Once again, due to the zero-range nature of the interaction, 
 after one single integration by parts on the 
last but one term, one has 
\begin{eqnarray}
\bigtriangleup  E^D_{II}&=&T\int d\bm x 
\left[ \frac{1}{16}\bm S(\bm x)\cdot\Delta \bm S(\bm x)-
\frac{1}{12}\bm S(\bm x)\cdot\bm T(\bm x)
+\frac{1}{12}
\stackrel{\leftrightarrow}{J}^2(\bm x)\right]; 
\end{eqnarray}
the square of the tensor of rank two $\stackrel{\leftrightarrow}{J}$ 
is defined, as usual, by 
$\stackrel{\leftrightarrow}{J}_q^2=
\sum_{\mu\nu} 
(J^{(q)}_{\mu\nu})^2$   
 and, for every  point $\bm x$, 
 the relation 
\begin{equation} 
\stackrel{\leftrightarrow}{J}_q^2=\underline J_q^2 +\frac{1}{2}\bm J_q^2+\frac{1}{3}(J^{(0)}_q)^2 
 \label{eqJJ}  
\end{equation} 
holds. 
By summing up 
 the two energies  $\bigtriangleup E^D_I$ and $\bigtriangleup  E^D_{II}$  
and taking into account the exchange, the total T-weighted 
contribution to the tensor energy 
 is  
\begin{eqnarray}
\bigtriangleup  E^{D+E}_T&=&\frac{T}{4}\int d\bm x 
\left\{
\left[ \frac{1}{4}\bm S(\bm x)\cdot\Delta \bm S(\bm x)-
\frac{1}{3}\bm S(\bm x)\cdot\bm T(\bm x)+
\frac{3}{4}(\bm \nabla\cdot \bm S(\bm x))^2+\bm S(\bm x)\cdot\bm F(\bm x)\right.\right.+\nonumber\\
&&\left.\left.-\frac{5}{9}{J^0}^2(\bm x)+\frac{5}{12}\bm J^2(\bm x)-\frac{1}{6}\underline{J}^2(\bm x)\right]+\right.\nonumber \\
&&\left. -\sum_{q=p,n}\left[ \frac{1}{4}\bm S_q(\bm x)\cdot\Delta\bm S_q(\bm x)-
\frac{1}{3}\bm S_q(\bm x)\cdot\bm T_q(\bm x)+
\frac{3}{4}(\bm\nabla \cdot\bm S_q(\bm x))^2+\bm S_q(\bm x)\cdot\bm F_q(\bm x)\right.\right.+\nonumber\\
&&\left.\left. -\frac{5}{9}(J^{(0)}_q)^2(\bm x)+\frac{5}{12}\bm J_q^2(\bm x)-\frac{1}{6}\underline{J}_q^2(\bm x)\right]\right\},
\label{fineT} 
\end{eqnarray}
where our equation 
\begin{equation}
-\int(\bm \nabla \cdot \bm S(\bm x))^2d\bm x=
2\int [\bm S(\bm x)\cdot\bm G(\bm x)+\bm S(\bm x)\cdot\bm F(\bm x)]d\bm x 
\label{SF_SG_eq}
\end{equation}
 has been used.
 As a matter of fact,   the  integration by parts of the square of the divergence of 
 the spin density   reads  (the integral will be omitted)
\begin{eqnarray}  
(\bm\nabla\cdot S(\bm x))^2&=&\sum_{\mu\nu}\nabla_{\mu}S_{\mu}(\bm x)\nabla_{\nu}S_{\nu}(\bm x)\nonumber\\
&=&-\sum_{\mu\nu}S_{\mu}(\bm x_2,\bm x'_2)\Big[\nabla_{1',\mu}\nabla_{1,\nu}S_{\nu}(\bm x_1,\bm x'_1)+
\nabla_{1,\mu}\nabla_{1,\nu}S_{\nu}(\bm x_1,\bm x'_1)+\nonumber\\
&&\nabla_{1',\mu}\nabla_{1',\nu}S_{\nu}(\bm x_1,\bm x'_1)+
\nabla_{1',\mu}\nabla_{1,\nu}S_{\nu}(\bm x_1,\bm x'_1)
\Big]_{\bm x_1=\bm x'_1=\bm x_2=\bm x'_2}   
\end{eqnarray} 
and one can recognize the structure of the $\bm S\cdot\bm F$ and $\bm S\cdot \bm G$ terms on the right hand side. 
 \\ 
\noindent 
Concerning the terms weighted 
by the U coefficient, 
 the first contribution reads 
\begin{eqnarray}
\bigtriangleup  E^D_{III}&=&
\frac{U}{8}
\sum_{i,j}\sum_{\omega_{1'},\omega_1,\omega_{2'},\omega_2}
 \sum_{\mu\nu}\int d\bm R 
 \Big\{\Big[
\Psi^{*(\mu)}_{i}(\bm x'_1, \omega'_1)\phi^*_{j}(\bm x'_2,\omega'_2)
-\phi_i(\bm r'_1, \omega'_1)\Psi^{*(\mu)}_{j}(\bm x'_2,\omega'_2)
\Big] 
\sigma_{\mu}^{(1)}\delta\sigma_{\nu}^{(2)}
\nonumber\\  
&&
\Big[
\Psi_i^{(\nu)}(\bm x_1,\omega_1)\phi_j(\bm x_2,\omega_2)-
\phi_i(\bm x_1,\omega_1)\Psi_j^{(\nu)}(\bm x_2,\omega_2)\Big]\Big\},
\end{eqnarray}
  where  the definition 
$\Psi_{\mu}(\bm y,\omega)=\left[\nabla_{\mu}\phi(\bm x,\omega)\right]_{\bm x=\bm y}$ 
and the short-hand notation $
\delta=
\delta(\bm x_1-\bm x_2)
\delta(\bm x_1-\bm x'_1)
\delta(\bm x_2-\bm x'_2)
$ 
have been introduced. 
By inserting the definition of the spin density, 
 after a few algebric 
 steps one  obtains 
\begin{eqnarray} 
\bigtriangleup  E^D_{III}&=& \frac{U}{8}\sum_{\mu\nu}\int d\bm R 
\Big\{ S_{\nu}(\bm x_2,\bm x_2')
\nabla_{1',\mu}\nabla_{1,\nu}
S_{\mu}(\bm x_1,\bm x_1')
-\nabla_{1',\mu}S_{\mu}(\bm x_1,\bm x_1')
\nabla_{2,\nu}S_{\nu}(\bm x_2,\bm x_2')+\nonumber\\
&&-\nabla_{1,\nu}S_{\mu}(\bm x_1,\bm x_1') 
 \nabla_{2',\mu}S_{\nu}(\bm x_2,\bm x_2')
+S_{\mu}(\bm x_1,\bm x_1')
\nabla_{2,\nu}\nabla_{2',\mu}
S_{\nu}(\bm x_2,\bm x_2')\Big\}_{\bm x_1=\bm x'_1=\bm x_2=\bm x'_2}\hspace{-2.cm}. 
\end{eqnarray} 
By using the relation~(\ref{1E})
 and the identity 
\begin{equation}
\sum_{\mu\nu}
S^{(q)}_{\mu}(\bm x_2,\bm x'_2)
\left(
\nabla'_{1,\nu}
\nabla_{1,\mu}+
\nabla_{1,\nu}
\nabla'_{1,\mu}\right)
S^{(q)}_{\nu}(\bm x_1,\bm x'_1)
=2
 \bm S_q(\bm x_2,\bm x'_2)\cdot
\bm F_q(\bm x_1,\bm x'_1),
\end{equation}
 one is led to  
\begin{eqnarray}
\bigtriangleup  E^D_{III}&=&\frac{U}{8}\int d\bm x_1 d\bm x_2 
 \Big\{2\bm S(\bm x_2)\cdot
\bm F(\bm x_1)
-\frac{1}{4}\Big[\bm \nabla_1 
\cdot\bm S(\bm x_1)\Big]
\Big[\bm \nabla_2 
\cdot\bm S(\bm x_2)\Big] 
-J_0(\bm x_1)
J_0(\bm x_2)\nonumber\\
&&-\frac{1}{4}\sum_{\mu\nu} 
\nabla_{1,\nu}
S_{\mu}(\bm x_1)\nabla_{2,\mu}
S_{\nu}(\bm x_2)
-\sum_{\mu\nu}
J_{\nu\mu}(\bm x_1)
J_{\mu\nu}(\bm x_2)\Big\}_{\bm x_1=\bm x_2}\hspace{-0.8cm}. 
\end{eqnarray}
By performing the same steps used to obtain Eq.~(\ref{first}) 
we get 
\begin{equation}
\bigtriangleup  E^D_{III}=\frac{U}{8}
\int d\bm x  
\left[2\bm S(\bm x)\cdot\bm F(\bm x)-\frac{1}{2}
(\bm \nabla \cdot\bm 
S(\bm x))^2 - {J_0}^2(\bm x) - 
 \sum_{\mu\nu}J_{\mu\nu}(\bm x)J_{\nu\mu}(\bm x)\right].
\end{equation}
On the basis of Eq.~(\ref{eqJ_}),    
one finally has 
\begin{equation}
\bigtriangleup  E^D_{III}=\frac{U}{8}
\int d\bm x \left[2\bm S(\bm x)\cdot\bm F(\bm x)-\frac{1}{2}(\bm \nabla \cdot\bm S(\bm x))^2 - 
\frac{4}{3}{J_0}^2(\bm x) 
+\frac{1}{2}\bm {J}^2(\bm x)
- 
\underline{J}^2(\bm x)
\right].  
\end{equation}
For the second U contribution of Eq.~(\ref{eqn}), 
the different spin-momentum structure allows a more compact expression 
\begin{eqnarray}
\bigtriangleup  E^D_{IV}&=&
-\frac{U}{12}
\sum_{i,j}\sum_{\omega_{1'},\omega_1,\omega_{2'},\omega_2}
\sum_{\mu} \int d\bm R \Big\{
\Psi^{*(\mu)}_{i}(\bm x'_1,\omega'_1)\phi^*_{j}(\bm x'_2,\omega'_2) 
 \nonumber\\
&&
\bm\sigma_1\cdot\bm\sigma_2\delta
\left[\Psi_i^{(\mu)}(\bm x_1,\omega_1)\phi_j(\bm x_2,\omega_2)
-
\phi_i(\bm x_1,\omega_1)\Psi^{(\mu)}_j(\bm x_2,\omega_2)
\right]\Big\}.
\end{eqnarray}
This is equal to  
\begin{eqnarray}
\bigtriangleup  E^D_{IV}&=&
-\frac{U}{12}\int  
 d\bm R \sum_{\mu\nu}
\left\{ S_{\nu}(\bm x_2,\bm x'_2)
{\nabla}_{1',\mu}
{\nabla}_{1,\mu}
S_{\nu}(\bm x_1,\bm x'_1)-
{\nabla}_{1',\mu}
S_{\nu}(\bm x_1,\bm x'_1)
{\nabla}_{2,\mu}
S_{\nu}(\bm x_2,\bm x'_2)
\right\}._{\bm x_1=\bm x'_1=\bm x_2=\bm x'_2}\hspace{-1.5cm}
\end{eqnarray}
By using the relations~(\ref{1E}) one more time  
and recognizing the scalar product between the spin and the kinetic density 
\begin{equation}
    \sum_{\mu} S^{(q)}_{\mu}(\bm x_2,\bm x'_2)
\bm\nabla_{1'}
\cdot\bm\nabla_{1}
S^{(q)}_{\mu}(\bm x'_1,\bm x_1)=
\bm S_q(\bm x_2,\bm x'_2)\cdot\bm T_q(\bm x_1,\bm x'_1)   
\end{equation}
one obtains 
\begin{eqnarray}
\bigtriangleup  E^D_{IV}&=&-\frac{U}{12}
\int d\bm x_1 d\bm x_2 
\Big\{\bm S(\bm x_2)\cdot
\bm T(\bm x_1)-
\frac{1}{4}\sum_{\mu\nu}
\nabla_{2,\mu}S_{\nu}(\bm x_2)
\nabla_{1,\mu}S_{\nu}(\bm x_1)
-\sum_{\mu\nu}
J_{\mu\nu}(\bm x_1)
J_{\mu\nu}(\bm x_2)
\Big\}_{\bm x_1=\bm x_2}\hspace{-0.8cm},\nonumber 
\end{eqnarray} 
 that is  
\begin{equation}
\bigtriangleup  E^D_{IV}=-\frac{U}{12}
\int d\bm x  \left[\bm S(\bm x)\cdot\bm T(\bm x)+
\frac{1}{4}\bm S(\bm x)\cdot\Delta \bm S(\bm x) 
-\stackrel{\leftrightarrow} J^2(\bm x)\right]. 
\end{equation} 
By adding together $\bigtriangleup  E^D_{III}$
and 
$\bigtriangleup  E^D_{IV}$, the final result is 
\begin{equation}
\bigtriangleup  E^D_{U}
=\int d\bm x \frac{U}{12}\left[3\bm S(\bm x)\cdot\bm F(\bm x)-
\bm S(\bm x)\cdot\bm T(\bm x)-\frac{1}{4}\bm S(\bm x)
\Delta
\bm S(\bm x)-\frac{3}{4}(\bm \nabla\cdot \bm S)^2(\bm x)-\frac{5}{3}{J_0}^2(\bm x)+\frac{5}{4}
\bm J^2(\bm x)-\frac{1}{2}\underline{J}^2(\bm x)
\right].  
\end{equation}
\\ The direct plus exchange provides  
 \begin{table}
\caption{Two possible combinations of the $D_{\alpha}$ coupling constants  
 for the direct functional's terms listed in the first row. The 
 set (a) corresponds to what is straightforwardly obtained for the direct contributions 
 through the derivation presented here, when Eq.~(\ref{SF_SG_eq})  is not used in the derivation.  
   The choice (b), on which the  result~(\ref{fineT}) is based,   
  leads to recover the expression of~\cite{Perl}.   
In both cases, 
  the Galilean invariance of the functional is fulfilled (see Eqs.~(\ref{GalGF})).  
\label{Tab9} }
\begin{ruledtabular}
\begin{tabular}{cccc}
set  &  $\bm S\cdot\bm G$ & $\bm S\cdot\bm F$ & $(\bm\nabla\cdot\bm S)^2$ \\
\hline 
(a) &$\frac{T}{4}$ & $\frac{U}{4}$ & $\frac{1}{16}(T-U)$ \\
(b) & 0 & $\frac{1}{4}(T+U)$ & $\frac{1}{16}(3T-U)$ \\
\end{tabular}
\end{ruledtabular}
\end{table}
\begin{eqnarray}
\bigtriangleup  E^{D+E}_U
&=&\frac{U}{12}\int d\bm x\left\{3\bm S(\bm x)\cdot\bm F(\bm x)-\bm S(\bm x)\cdot\bm T(\bm x)-\frac{1}{4}\bm S(\bm x)
\cdot\Delta
\bm S(\bm x)-\frac{3}{4}(\bm \nabla \cdot\bm S)^2(\bm x)+\right.\nonumber\\
&&\left.-\frac{5}{3}{J_0}^2(\bm x) 
+\frac{5}{4}\bm J^2(\bm x)-\frac{1}{2}\underline{J}^2(\bm x)
\right.
\nonumber\\
&&\left.
+\sum_{q=p,n}\left[3\bm S_q(\bm x)\cdot\bm F_q(\bm x)-\bm S_q(\bm x)\cdot\bm T_q(\bm x)-
\frac{1}{4}
\bm S_q(\bm x)\cdot 
\Delta
\bm S_q(\bm x)-\frac{3}{4}(\bm \nabla\cdot \bm S_q)^2(\bm x)\right.\right.\nonumber\\
&&\left.\left.-\frac{5}{3}(J^{(0)}_{q})^2(\bm x)+\frac{5}{4}\bm J_q^2(\bm x)-\frac{1}{2}\underline{J}_q^2(\bm x)
\right]\right\}.\nonumber\\
&&\label{fineU}
\end{eqnarray}
 The sum of the integrand in Eqs.~(\ref{fineT}) and~(\ref{fineU}) provides the tensor 
  contribution to the standard Skyrme energy density functional, denoted by 
 $\mathcal{H}_{tens}$ in Eq.~(\ref{Htot}). \\ 
 By defining as $D_{\alpha}$  ($E_{\alpha}$) 
 the weight of a generic 
 term  $\alpha$  belonging to  the direct (exchange) contributions to the energy, the relations  
   $D_{\alpha}$=$C^{\alpha}_0$-$C^{\alpha}_1$ and 
    $D_{\alpha}+E_{\alpha}$=$C^{\alpha}_0$+$C^{\alpha}_1$ hold, where 
  $C_{T=0,1}^{\alpha}$ are  the isoscalar and isovector coupling constants 
 employed in some other works. 
   By rearranging the result  in terms of the isoscalar and isovector densities,  
the sum of the expressions~(49a)+(49b)  of~\cite{Perl}  is recovered. Otherwise,  
 when the $\bm S\cdot\bm G$ terms from~(\ref{first}) are retained, 
  the $\bm S\cdot\bm F$ and 
  $(\bm\nabla\cdot\bm S)^2$ coupling constants in~(\ref{fineT})     
  change 
 according to the combination (a) of Tab.~\ref{Tab9};  in such  a way,  
  one  obtains, in the proton-neutron formalism, the more general 
   formulation~(\ref{Htot}), where $A_{\alpha}=D_{\alpha}+E_{\alpha}$ and  
 $B_{\alpha}=D_{\alpha}$. Since $E_{\alpha}=\mp D_{\alpha}$ for the T (-) and U (+) contribution, 
   only the U-weigthed  terms 
  contribute to $A_{\alpha}$.       
\end{widetext}

\section{Densities and currents of the Skyrme-tensor EDF}
\label{App2}
In this Appendix,  the expressions of the 
 densities and currents entering the Skyrme energy density functional~(\ref{Htot}) 
 are provided. 
For the sake of simplicity, we take as implicit 
    the sum over the 
 wave-functions index, which includes all the quantum numbers 
 with the exception of the spin projection $\omega$, which is separately written.  
\\\indent 
When boosting the Hartree-Fock single-particles, they gain an imaginary part. 
 We  adopt the  definitions of Tab.~\ref{Tab10},
 where $\phi_{\omega}=\phi(\bm x,\omega)$, 
$k=x,y,z$ and 
$\omega=2m_s=\pm 1$,   
 and of Tab.~\ref{Tab11},  where the second order terms are listed. 
   \begin{table}
\caption{Conventions (I) adopted in the current Appendix. 
\label{Tab10} }
\begin{ruledtabular}
\begin{tabular}{cc}
$\phi^R_{\omega}$&$\Re\phi_{\omega}$\\
$\phi^I_{\omega}$&$\Im\phi_{\omega}$\\
$\Psi^{\omega}_k$&$\bigtriangledown_k\phi_{\omega}$\\
$\tilde\Psi^{\omega}_k$&$\bigtriangledown_k\phi^*_{\omega}$\\
$\bm\Psi_{\omega}$&$\bm{\nabla}\phi_{\omega}$\\       
$\tilde{\bm{\Psi}}_{\omega}$&$\bm{\nabla}\phi^*_{\omega}$\\
$\Psi^{R,\omega}_k$&$\Re\Psi_k^{\omega}$\\
$\Psi^{I,\omega}_k$&$\Im\Psi_k^{\omega}$\\
$\bm\Psi^{R(I)}_{\omega}$&$\Re(\Im)\bm\Psi_{\omega}$ 
\end{tabular}
\end{ruledtabular}
\end{table}
  \begin{table}
\caption{Conventions (II) adopted in the current Appendix. 
\label{Tab11} }
\begin{ruledtabular}
\begin{tabular}{cc}
$ \Psi^{\omega}_{kk'}$&$\nabla_k\nabla_{k'}\phi_{\omega}$\\
$\tilde\Psi^{\omega}_{kk'}$&$\nabla_k\nabla_{k'}\phi^*_{\omega}$\\
$\Psi^{R,\omega}_{kk'}$&$\Re\Psi_{k,k'}^{\omega}$\\
$\Psi^{I,\omega}_{kk'}$&$\Im\Psi_{k,k'}^{\omega} $
\end{tabular}
\end{ruledtabular}
\end{table}
In the following, 
 all the relations 
 hold if considering 
 isoscalar, isovector, neutron or proton densities.   
\\\indent 
 The   densities which are even under the time-reversal operation $T_K:\phi_i(\bm x,\omega)\rightarrow -\omega\phi^*(\bm x,-\omega)$ 
 ($T_K=-i\sigma_yK_0$,  where $K_0$ denotes the complex-conjugation),  
    include 
\begin{itemize}
\item{the particle density
\begin{eqnarray}
\rho(\bm x)&=&\sum_{\omega}\phi^*(\bm x', \omega)\phi(\bm x, \omega)|_{\bm{ x}=\bm{x}'}\\
&=&\sum_{\omega}[(\phi^R_{\omega})^2+(\phi^I_{\omega})^2]=\sum_{\omega}|\phi_{\omega}|^2;
\end{eqnarray}
}
\item{the kinetic density
\begin{eqnarray}
\tau(\bm x)&=&[\bm\nabla\cdot\bm\nabla'\rho(\bm x,\bm x')]_{\bm x=\bm x'}\\
 &=&\sum_{\omega}[(\bm\nabla\phi^R_{\omega})^2+(\bm\nabla\phi^I_{\omega})^2]
=\sum_{\omega}|\bm\Psi_{\omega}|^2; 
\end{eqnarray}
}
\item{the spin-current pseudo-tensor $\stackrel{\leftrightarrow}J$
\begin{equation}
\stackrel{\leftrightarrow}J(\bm{x})=\frac{1}{2i}\left[(\bm\nabla-\bm\nabla')\otimes\bm{S}(\bm{x},\bm{x}')\right]_
{\bm{x}=\bm{x}'}.  
\end{equation}
 The nine components ($k=x,y,z$) can be implemented as  
\begin{eqnarray}
J_{k x}&=&\hspace{-0.5cm}
\sum_{\omega,\omega'=\pm 1,\omega\ne\omega'}\hspace{-0.5cm}
[\phi^R_{\omega}\Psi_k^{I,\omega'}-\phi^I_{\omega}\Psi_k^{R,\omega'}]\nonumber\\
&=&+\Im\left[\phi^*_+\Psi_k^- +\phi^*_-\Psi_k^+\right]\\
J_{k y}&=&\sum_{L=R,I}[\phi^L_-\Psi_k^{L,+}-\phi^L_+\Psi_k^{L,-}]\nonumber\\
&=&-\Re\left[\phi^*_+\Psi_k^- -\phi^*_-\Psi_k^+\right]\\
J_{k z}&=&\phi^R_+\Psi_k^{I,+}-\phi^I_+\Psi_k^{R,+}+
\phi^I_-\Psi_k^{R,-}-\phi^R_-\Psi_k^{I,-}\nonumber\\
&=&+\Im\left[\phi^*_+\Psi_k^+ -\phi^*_-\Psi_k^-\right]. 
\end{eqnarray}}
\end{itemize}
The  time-odd densities are 
\begin{itemize}
\item{the spin density
\begin{equation}
\bm{S}(\bm x)=\sum_{\omega,\omega '} 
  \left. \phi^{*
}(\bm{x} ',\omega')
\phi(\bm{ x},\omega) 
\langle 
\omega '|\bm{ \sigma}|\omega \rangle \right |_{\bm{ x}=\bm{x}'},
\end{equation}
where each component corresponds to  
\begin{eqnarray}
S_x&=&2[\phi^R_+\phi^R_-+\phi^I_+\phi^I_-]
=2\Re[\phi^*_+\phi_-]\\
S_y&=&2[\phi^R_+\phi^I_--\phi^R_-\phi^I_+]
=2\Im[\phi^*_+\phi_-]\\
S_z&=&\hspace{-0.3cm}\sum_{L=R,I}\hspace{-0.1cm}[\phi^L_+\phi^L_+-\phi^L_-\phi^L_-]=\Re[\phi^*_+\phi_+-\phi^*_-\phi_-]
\end{eqnarray}
}
\item{the momentum density 
\begin{equation}
\bm j(\bm x)=-\frac{i}{2}[(\bm\nabla-\bm\nabla')\rho(\bm x,\bm x')]_{\bm x=\bm x'}, 
\end{equation}
 with components equal to 
\begin{eqnarray}
j_k&=&\sum_{\omega=\pm}[\phi^R_{\omega}\Psi_x^{I,\omega}-\phi_{\omega}^I\Psi_x^{R,\omega}]\nonumber\\
&=&\Im[\phi^*_+\Psi_x^++\phi^*_-\Psi_x^-];
\end{eqnarray}}
\item{the spin kinetic density 
\begin{equation}
 \bm T(\bm x)=\left[\bm\nabla\cdot\bm\nabla '\bm S(\bm x,\bm x')\right]_{\bm x=\bm x'}, 
\end{equation}
where
\begin{eqnarray}
T_x&=&2[\bm\Psi^R_+\bm\Psi^R_-+\bm\Psi^I_+\bm\Psi^I_-]\nonumber\\
&=&2\Re[\bm\Psi_-\tilde{\bm\Psi}_+]\\
T_y&=&2[\bm\Psi^R_+\bm\Psi^I_--\bm\Psi^I_+\bm\Psi^R_-]\nonumber\\
&=&2\Im[\bm\Psi_-\tilde{\bm\Psi}_+]\\
T_z&=&\hspace{-0.2cm}\sum_{L=R,I}[\bm\Psi_+^L\bm\Psi_+^L-\bm\Psi_-^L\bm\Psi^L_-]\nonumber\\
 &=&\Re[\bm\Psi_+\tilde{\bm\Psi}_+-\bm\Psi_-\tilde{\bm\Psi}_-];
\end{eqnarray}}
\item{the pseudo-vector $\bm F(\bm x)$ density   
\begin{equation}
\hspace{0.8cm}\bm F(\bm x)=\frac{1}{2}\left \{\left[ (\bm \nabla '\otimes \bm \nabla )  +(\bm \nabla \otimes \bm \nabla '    )\right]
   \bm S(\bm x,\bm x')\right\}_{\bm x=\bm x'}. 
\nonumber\end{equation}
 It can be conveniently represented by splitting
   each $k$  component 
 in three contributions 
\begin{equation}
F_{k}(\bm x)=\sum_{k'=x,y,z}F_{k}^{(k')}(\bm x),\\
\end{equation}
where the single terms read  
\begin{eqnarray}
F_{x}^{(x)}&=&2[\Psi_x^{R,+}\Psi_x^{R,-}+\Psi_x^{I,+}\Psi_x^{I,-}] \nonumber\\
&=&2\Re[\tilde\Psi_x^+\Psi_x^-]   \\
F_{x}^{(y)}&=&\hspace{-0.5cm}
\sum_{k,k'=x,y;k\ne k'}\hspace{-0.5cm}[\Psi_k^{I,-}\Psi_{k'}^{R,+}-\Psi_{k'}^{I,+}\Psi_{k}^{R,-}] 
\nonumber\\
&=& \Im[\tilde\Psi_x^+\Psi^-_y-\tilde\Psi_x^-\Psi_y^+]    \\
F_{x}^{(z)}&=& \sum_{L=R,I}[\Psi_{x}^{L,+}\Psi_{z}^{L,+}-\Psi_{x}^{L,-}\Psi_{z}^{L,-}]\nonumber\\
&=&\Re[\tilde\Psi_x^+\Psi_z^+-\tilde\Psi^-_x\Psi_z^-]     \\
&&\nonumber\\
&&\nonumber\\
F_{y}^{(x)}
&=& \sum_{L=R,I} [\Psi_{y}^{L,+}\Psi_{x}^{L,-}+\Psi_{y}^{L,-}\Psi_{x}^{L,+}]\nonumber\\
&=&\Re[\tilde\Psi_y^+\Psi_x^-+\tilde\Psi_y^{-}\Psi_x^+]\\ 
F_{y}^{(y)}
&=&2[\Psi_y^{I,-}\Psi_y^{R,+}-\Psi_y^{R,-}\Psi_y^{I,+}]\nonumber\\
&=&-2\Im[\tilde\Psi_y^-\Psi_y^+]\\
F_{y}^{(z)}&=&F_{x}^{(z)}\big |_{x\rightarrow y}\\
&&\nonumber\\
&&\nonumber\\
F_{z}^{(x)}
&=&F_{y}^{(x)}\big |_{y\rightarrow z}\\
F_{z}^{(y)}
&=&F_{x}^{(y)}\big |_{x\rightarrow z}\\
F_{z}^{(z)}
&=&
\sum_{L=R,I}[(\Psi_{z}^{L,+})^2-(\Psi_{z}^{L,-})^2]\nonumber\\
&=&\Re[\tilde\Psi_z^{+}\Psi_z^+-\tilde\Psi_z^-\Psi_z^-]; 
\end{eqnarray}
}
\item{
the pseudo-vector $\bm G(\bm x)$ density
\begin{equation}
\hspace{0.8cm}\bm G(\bm x)=\frac{1}{2}\left \{\left[ (\bm \nabla '\otimes \bm \nabla'   )+(\bm \nabla \otimes \bm \nabla    )\right]
   \bm S(\bm x,\bm x') \right\}_{\bm x=\bm x'}; 
\nonumber\end{equation}
 by proceeding similarly to $\bm F(\bm x)$, one obtains  
\begin{eqnarray}
G_x^{(x)}&=&\sum_{L=R,I}[\phi^{L}_+\bigtriangleup_x\phi^{L}_-
+\phi^{L}_-\bigtriangleup_x\phi^{L}_+] \nonumber\\
&=&\Re[\phi^*_+\bigtriangleup_x\phi_-+
\phi^*_-\bigtriangleup_x\phi_+]\\
G_x^{(y)}&=&\phi^{R}_+\Psi_{xy}^{I,-}-\phi^{I}_+\Psi_{xy}^{R,-}
+\phi^{I}_-\Psi_{xy}^{R,+}-\phi^{R}_-\Psi_{xy}^{I,+}\nonumber\\
&=&\Im[\phi^*_+\Psi_{xy}^--\phi^*_-\Psi_{xy}^+]\\
G_x^{(z)}&=&\sum_{L=R,I}[\phi^{L}_+\Psi_{xz}^{L,+}-\phi^{L}_-\Psi_{xz}^{L,-}]\nonumber\\
&=&\Re[\phi^{*}_+\Psi_{xz}^{+}-\phi^*_{-}\Psi_{xz}^{-}]\\
&&\nonumber\\
&&\nonumber\\
G_y^{(x)}&=&\sum_{L=R,I}[\phi^{L}_+\Psi_{yx}^{L,-}+\phi^{L}_-\Psi_{yx}^{L,+}] \nonumber\\
&=&\Re[\phi^*_+\Psi_{yx}^-+\phi^*_{-}\Psi_{yx}^{+}]\\
G_y^{(y)}&=&\phi^{R}_-\bigtriangleup_y\phi^{I}_--\phi^{I}_+\bigtriangleup_y\phi^{R}_-+\nonumber\\
&&\phi^{I}_-\bigtriangleup_y\phi^{R}_+-
\phi^{R}_-\bigtriangleup_y\phi^{I}_+\nonumber\\
&=&\Im[\phi^*_+\bigtriangleup_y\phi_{-}-\phi^*_-\bigtriangleup_y\phi_{+}]\\
G_y^{(z)}&=&G_x^{(z)}\big|_{x\rightarrow y}\\
&&\nonumber\\
&&\nonumber\\
G_z^{(x)}&=&G_y^{(x)}\big|_{y\rightarrow z}\\
G_z^{(y)}&=&G_x^{(y)}\big|_{x\rightarrow z}\\
G_z^{(z)}&=&\sum_{L=R,I}[\phi^{L}_+\bigtriangleup_{z}\phi^{L}_+-\phi^{L}_-\bigtriangleup_{z}
\phi^{L}_-]\nonumber\\
&=&\Re[\phi^*_{+}\bigtriangleup_{z}\phi_+-\phi^*_{-}\bigtriangleup_{z}\phi_-].
\end{eqnarray}} 
\end{itemize}

\end{document}